\documentclass[openacc]{rstransa}

\usepackage{todonotes}
\usepackage{multirow}
\usepackage{graphicx}

\begin{document}

\title{Neural Network Design for Energy-Autonomous AI Applications using Temporal Encoding}

\author{
Sergey Mileiko$^{1}$, Thanasin Bunnam$^{2}$, Fei Xia$^{3}$, Rishad Shafik$^{4}$, Alex Yakovlev$^{5}$ and Shidhartha Das$^{6}$}

\address{$^{1}$S.Mileiko2@newcastle.ac.uk\\
$^{2}$T.Bunnam2@newcastle.ac.uk\\
$^{3}$Fei.Xia@newcastle.ac.uk\\
$^{4}$Rishad.Shafik@newcastle.ac.uk\\
$^{5}$Alex.Yakovlev@newcastle.ac.uk\\
$^{6}$Shidhartha.Das@arm.com}

\subject{Edge computing, artificial intelligence.}

\keywords{Neural networks, hardware design, energy efficiency, energy autonomy.}

\corres{Serhii Mileiko\\
\email{S.Mileiko2@newcastle.ac.uk}}

\begin{abstract}
Neural Networks (NNs) are steering a new generation of artificial intelligence (AI) applications at the micro-edge. Examples include wireless sensors, wearables and cybernetic systems that collect data and process them to support real-world decisions and controls. For energy autonomy, these applications are typically powered by energy harvesters. As harvesters and other power sources which provide energy autonomy inevitably have power variations, the circuits need to robustly operate over a dynamic power envelope. In other words, the NN hardware needs to be able to function correctly under unpredictable and variable supply voltages.

In this paper, we propose a novel NN design approach using the principle of pulse width modulation (PWM). PWM signals represent information with their duty cycle values which may be made independent of the voltages and frequencies of the carrier signals. We design a PWM-based perceptron which can serve as the fundamental building block for NNs, by using an entirely new method of realising arithmetic in the PWM domain. We analyse the proposed approach building from a $3 \times 3$ perceptron circuit to a complex multi-layer NN. Using handwritten character recognition as an exemplar of AI applications, we demonstrate the power elasticity, resilience and efficiency of the proposed NN design in the presence of functional and parametric variations including large voltage variations in the power supply. 
\end{abstract}

\maketitle

\section{Introduction}

Advances in sensing devices are causing a shift towards the fourth industrial revolution~\cite{schwab2017fourth}. 
The large volumes of the data produced by these devices are enabling a new generation of artificial intelligence (AI) systems at the micro-edge that are designed to infer important decisions in the real world~\cite{javaid2018intelligence}. 
A promising direction of these AI Systems is the leap towards perpetual computability, allowing always available local AI service. To enable this, designers of pervasive AI system are facing two grand challenges: \textit{energy efficiency} and \textit{energy autonomy}~\cite{biswas2018convram, chen2019lpwan, sharbati2018lowpower,neftci2018data}. 

Energy efficiency refers to economising the energy consumption of elementary compute operations. The aim is to prolong operating lifetime with a given energy budget, typically defined by the batteries. Reducing energy requires careful design considerations at device-, circuit- and system levels. Examples include reducing device geometry~\cite{ShafikPower}, scaling operating voltage~\cite{Shafik2016a} and designing circuits with reduced or approximate logic~\cite{Qiqieh2017}. 

New generations of pervasive AI-based systems require maintenance-free long-life. As such traditional energy-efficient design principles applied in battery-operated systems are not feasible, as they need periodic re-charging and replacements. Portable energy harvesters, which produce electrical energy to supply to computation loads by scavenging energy from the environment, are gradually making inroads. Such a scheme of energy harvesting can remove the need of maintenance in favour of energy autonomy. However, mitigating their energy variations needs computational capability over a dynamic power envelope, otherwise known as power elasticity~\cite{8330023,beeby2010energy}. 

Despite advances in low-power design methodologies, the energy footprint of existing AI systems, such as Neural Networks (NNs), has generally remained high~\cite{yang2017designing}. Our persistence in using arithmetic-heavy circuits with growing algorithmic complexities is a major contributor to this. For instance, object detection using deep NNs may require a hundred to over ten thousand times the energy needed by the traditional histogram of oriented gradient techniques~\cite{suleiman2017towards}. Due to such poor efficiency, the widespread adoption of energy-autonomous AI hardware at the micro-edge has proven challenging~\cite{Luca2018}.

To appreciate the importance of efficient AI hardware design, we show the example of a perceptron, whose idea originates from Rosenblatt's work of 1958~\cite{Rosenblatt58theperceptron:}. It is a basic building block of NNs used in AI applications~\cite{Hagan:1997:NND:249049,wilson1994multi,adeli1989perceptron}. It consists of an input vector, a set of weights and a bias to produce binary classification outcomes, as follows:
\begin{equation}
f(x) = \begin{cases}
      1, & \text{if}\ \textbf{w}.\textbf{x} + b > 0 \\
      0, & \text{otherwise}
    \end{cases}
\end{equation}
where $w$ is a vector of real-valued weights, \textbf{w}.\textbf{x}  is the dot product $\sum_{i=1}^{m}w_i x_i $ with $m$ number of inputs, and $b$ is the bias. The process of deciding the appropriate weights (\textbf{w}), often also known as training, serves as the basic principle of supervised learning. When $m$ becomes large, it approximates the behaviour of a biological neuron.
\begin{figure}[ht]
    \centering
    \includegraphics[scale=0.7]{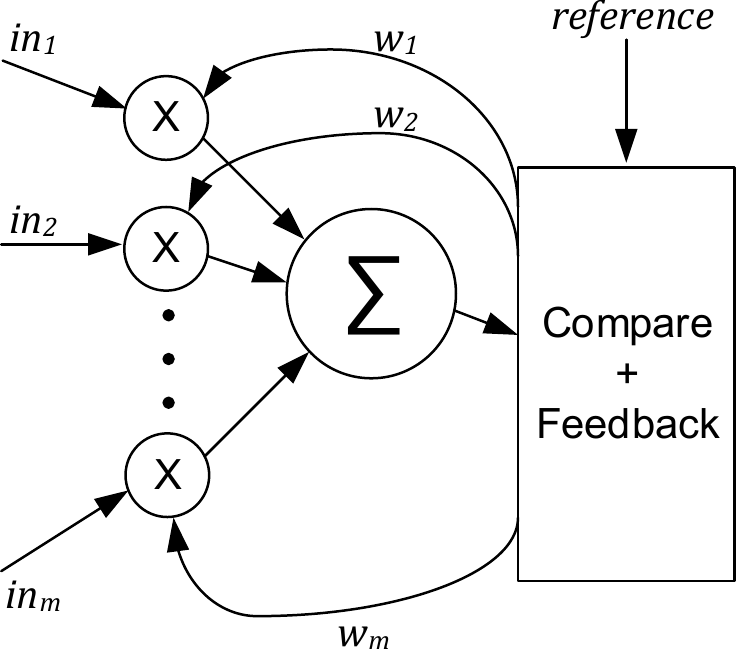}
    \caption{Structural organisation of a perceptron, which is the basic building block of NNs.}
    \label{fig:perc_struct_intro}
\end{figure}
Figure~\ref{fig:perc_struct_intro} shows the typical structure of a perceptron~\cite{hung1991model,jeong1994design}. At its core is an adder that sums $m$ weighted inputs. The result of the addition is compared with a reference during the training phase, during which the weights are updated to ensure the reference is matched. For hardware implementation, multiplication and addition are crucial arithmetic circuits in a perceptron~\cite{1243906}. Such arithmetic operations require significant area and power costs, which depend on the number of input-weight pairs, the precision of the multipliers\slash adders, their underlying technology nodes and algorithmic complexities.

Over the years, substantial research has been dedicated to improving the energy efficiency of AI hardware~\cite{chen2018understanding}. A vast body of this research has predominantly remained within the remits of Landauer's logic boundaries for energy or power reduction~\cite{keyes1970minimal}. Reducing threshold voltage that defines the logic boundaries and designing new low-complexity architectures are key to achieving this. Andri \textit{et al.}~\cite{Luca2018} proposed a NN architecture that showed how high-performance NN operations can be achieved by parallel logic blocks. These blocks are designed using low-threshold technology nodes that are faster and ultra-low power. Prado \textit{et al.}~\cite{de2018quenn} showed a logic approximation method applied in parallel NNs. Due to low-complexity architecture the individual components are faster and more energy-efficient. Among others, Qiqieh \textit{et al.}~\cite{Qiqieh2017} proposed logic compression approaches for reducing power consumption, area and critical path delay of NNs. By combining the circuit-level approaches with online system-wide techniques, significant energy reduction was reported.

However, reducing power or energy alone using the above principles, does not solve the problem of energy-autonomous pervasive AI systems~\cite{8330023}. These systems will need to be able to not only work with limited power supplies but also survive extreme variations as power regulation and energy storage options are limited and expensive in low-end micro-edge devices~\cite{Yak2011,8330023}. Indeed, these systems will need to be built with natural power elasticity to operate over a large power domain~\cite{shang2014,Yak2011,8330023}. 

Existing perceptron designs are predominantly digital, although a number of analogue implementations have been reported~\cite{7551398}~\cite{isscc_2016_chen_eyeriss}. The digital designs can operate over a range of powers defined by paired voltages ($V_{dd}$) and frequencies ($f$). These designs are however vulnerable to dynamic power supply variations, for example conditions where voltage of the power source changes in time and continuous $V_{dd}$ and $f$ pairing can prove expensive under limited energy budgets. As such, existing designs have poor power elasticity that prevents them from providing useful computation under unreliable or unstable power supply conditions.

In this paper, for the first time, we tackle the power-elastic AI hardware design. Our design underpins a radically new approach of duty cycle based computing for pulse-width modulated (PWM) signals using a number of parallel inverters. These  building blocks are then integrated as part of a higher level analysis method to support system-wide investigations in the context of an exemplar application. Our key motivation to use the duty-cycle time-domain representation of data is due to its potential fundamental resilience to dynamic variations in the amplitude and frequency of the signal, which are inevitable for energy autonomous systems drawing energy from the environment. The other motivating factor is the natural ability of CMOS logic to perform multiplication and addition operation on the duty-cycled inputs. This is enabled by the inherent effects of proportionally ratioed current switching in CMOS networks between P and N subnets during the operational cycle of each gate. This gives way to implementing the PWM-based compute functions directly on CMOS logic gates. Thus, one of the important goals we pursue in this investigation is to verify our \textbf{hypothesis} that the combined use of PWM representation and CMOS logic, with minimal use of additional analogue (ideally, only passive components) electronics, will deliver the sought efficiency and robustness of AI hardware.

In more concrete terms, the aim of this paper is to design and demonstrate a voltage and frequency elastic perceptron, which performs its arithmetic computation in the PWM-coded format for robustness to energy supply variations. To this purpose, a method of modelling and analysing such power and frequency elastic NN components also needs to be developed. The main \textbf{\textit{contributions}} are:
\begin{enumerate}
    \item a mixed-signal perceptron design using duty cycle-based temporal weight encoding and input switching via inverters, 
    \item extensive validation experiments in Cadence Analog Design tool demonstrating the perceptron design's resilience in the presence of static or dynamic voltage and frequency variations,
    \item a mathematical model describing the input-output relations of the proposed perceptron for system-level design and analysis to support its use in NN design, and
    \item configurations, analyses and evaluations using an example PWM perceptron-based NN which solves the MNIST handwritten digits classification.
\end{enumerate}

This rest of the paper is organised as follows: Section~\ref{sec:circuit_design} establishes the method of designing the PWM based perceptron: the idea of the PWM to voltage conversion, performing the arithmetic operations, and the implementation of the voltage to PWM converter. Section~\ref{sec:results} validates the approach using a number of parametric sweeps to demonstrate the frequency elasticity and power resilience. Section~\ref{sec:discussions} discusses the strengths and weaknesses of the proposed approach, and the ideas of the future improvements. Finally,  Section~\ref{sec:concl} concludes the paper.

\section{Method}
\label{sec:circuit_design}

This section focuses on the design of the PWM-based perceptron, including the fundamental theories, the circuits of its constituent parts, methods of PWM-based arithmetic, leading to the construction of NNs. The design methods form the basis of extensive analysis supporting the validation of the perceptrons integrated in a NN. 

A perceptron capable of voltage and frequency elasticity may be constructed by exploiting the fact that relative temporal properties, such as a PWM's duty cycle, are resilient to voltage and frequency variations. As the supply voltage reduces, any oscillatory activity, such as a clock signal, may show a reduced amplitude and reduced frequency. However, the ratio between the time within a period when the clock signal is high and the time within a period when the clock signal is low stays the same as both would increase at the same rate. 

Our method, therefore, is dedicated to finding ways of exploiting this fact by transferring computation from the digital domain, which is affected by voltage and frequency variations, to the relative temporal domain, which is not. This means making use of PWM-based techniques.

\subsection {Principles of Duty Cycle to Voltage Conversion}

\begin{figure}[ht]
    \centering
    \begin{minipage}[b]{0.45\linewidth}
        \centering
        \includegraphics[width=\textwidth]{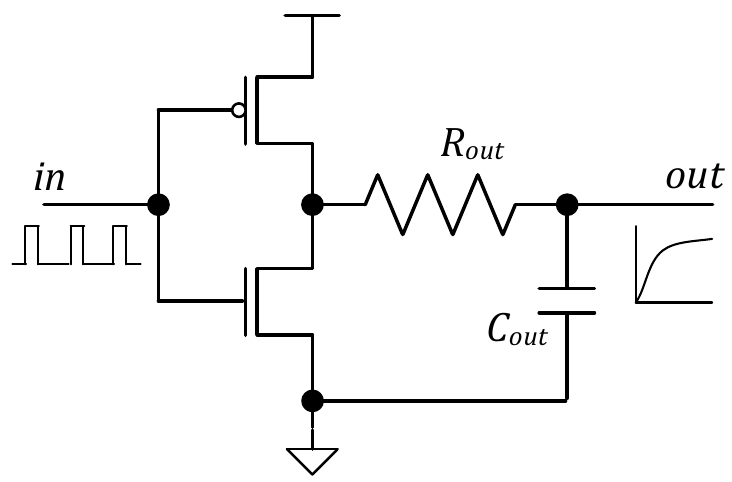}
        \caption{A CMOS-based inverter circuit.}
        \label{fig:inverter}
    \end{minipage}
    \hspace{0.5cm}
    \begin{minipage}[b]{0.45\linewidth}
        \centering
        \includegraphics[width=\textwidth]{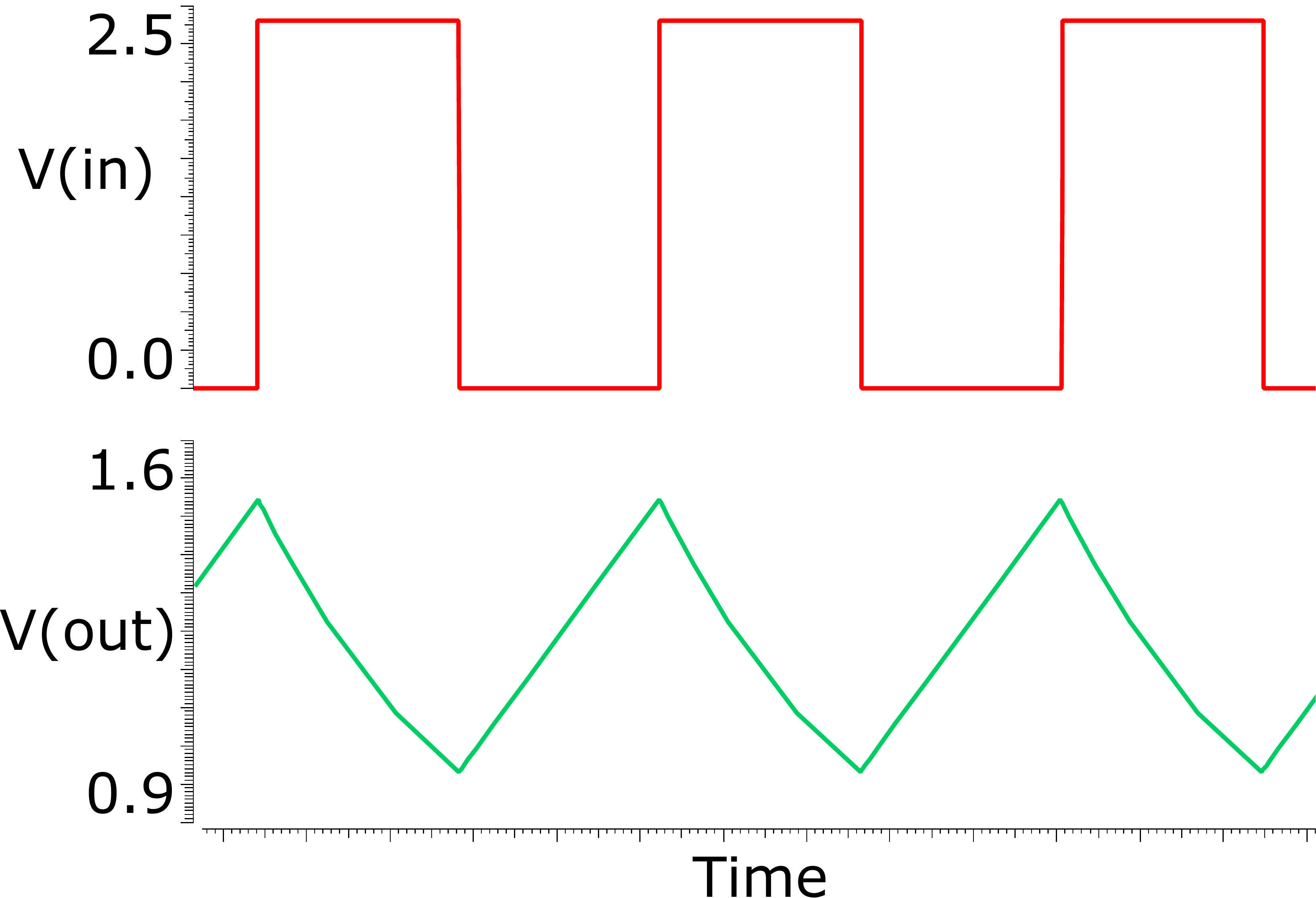}
        \caption{Inverter output with PWM-coded input.}
        \label{fig:inv_inout}
    \end{minipage}
\end{figure}

Figure~\ref{fig:inverter} shows an inverter-based PWM to voltage converter, which produces an output voltage whose value represents the value carried by the input PWM signal, i.e. its duty cycle. Here we exploit the principle that if the input of an inverter is a periodic signal, such as a clock, the average voltage on its output is inversely proportional to the duty cycle of the input signal. In other words, the analogue average value of the inverter's output voltage encodes the value of the duty cycle of the input signal. Since an inverter is a digital component, whose output equals to logic $'0'$ or $'1'$ at any moment in time, it needs to be "analogised" (i.e. transcoded) in order to convert the input duty cycle into the output voltage that is a corresponding proportion of the supply voltage. This may be achieved by the following ways:

\begin{itemize}

\item increasing the input switching frequency,

\item increasing the output capacitance,

\item limiting the output current.

\end{itemize}

For the inverter-based PWM to voltage converter shown in Figure~\ref{fig:inverter}, with the input clock duty cycle at $50\%$, the average output voltage is around $V_{dd}/2$~(Figure~\ref{fig:inv_inout}). This is due to the fact that during the interval of time when the input is low the output capacitance is charged with current from the power source via the PMOS transistor, and during the interval of input being high the capacitance is discharged via the NMOS transistor. With a $50\%$ duty cycle these two periods of time are the same length and, assuming the transistors are balanced, their voltages average out to half the supply voltage. When the duty cycle deviates from $50\%$ the average value of the output voltage deviates from $V_{dd}/2$ proportionally in the same direction. 

\begin{figure}[ht]
    \centering
    \begin{minipage}[b]{0.45\linewidth}
        \centering
        \includegraphics[width=0.6\textwidth]{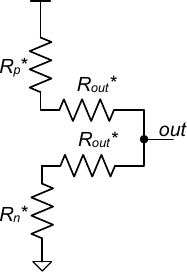}
        \caption{PWM inverter equivalent circuit, approximated as a voltage divider.}
        \label{fig:inv_eq}
    \end{minipage}
    \hspace{0.5cm}
    \begin{minipage}[b]{0.45\linewidth}
        \centering
        \includegraphics[width=\textwidth]{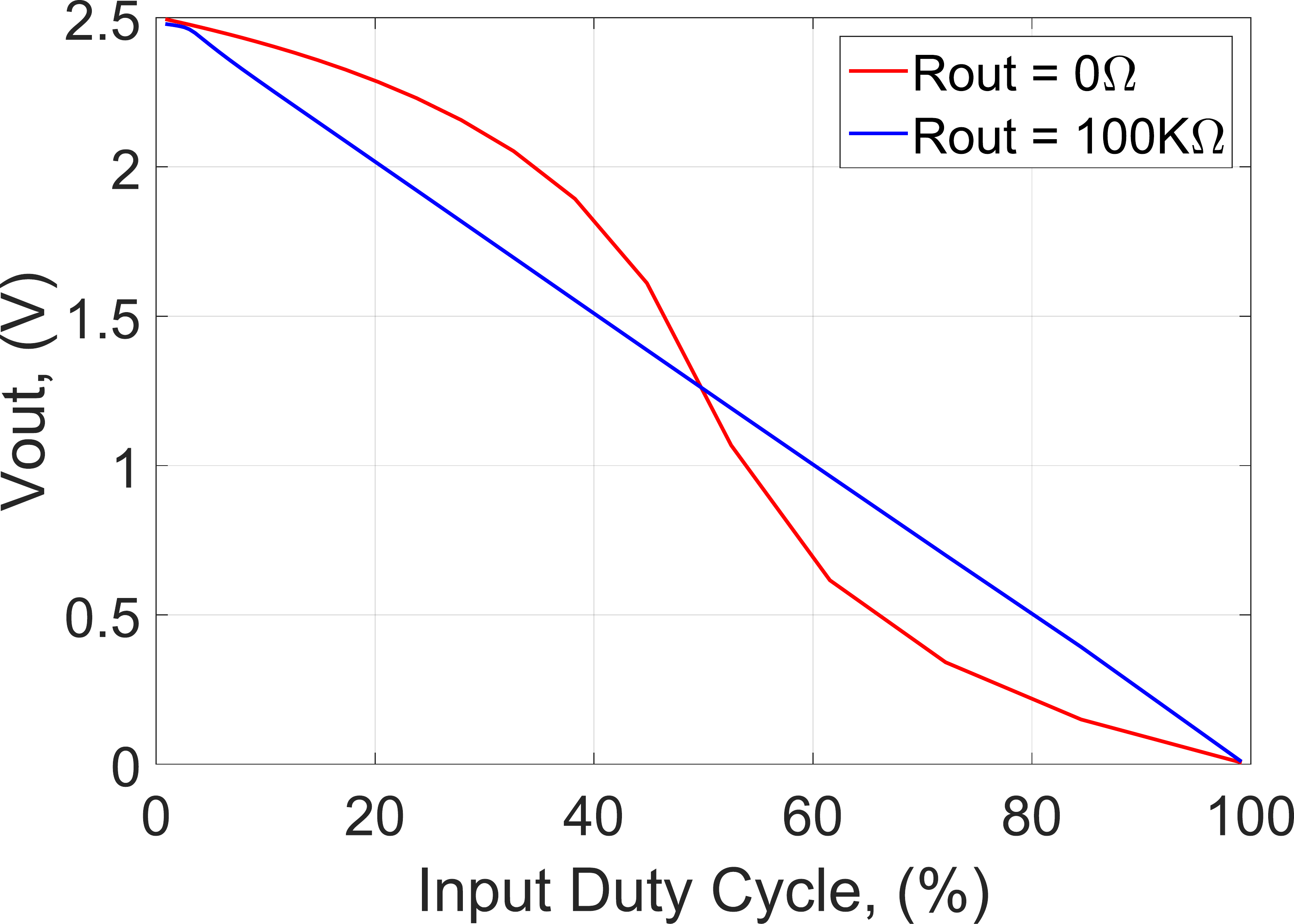}
        \caption{Output voltage of the PWM inverter vs input duty cycle.}
        \label{fig:inv_vout_dc}
    \end{minipage}
\end{figure}

If the frequency is high enough that the output capacitor is never fully charged or discharged, the inverter may be equivalently represented as a resistive voltage divider~(see Figure~\ref{fig:inv_eq}). The output voltage of such a divider can be calculated using the following equation.
\begin{equation}
\label{eq:voltage_divider}
	V_{out} = (V_{dd} - GND) \cdot \frac{R_{n}^{*}+R_{out}^{*}}{(R_{n}^{*}+R_{out}^{*})+(R_{p}^{*}+R_{out}^{*})}.
\end{equation}
where $R_{n}$ and $R_{p}$ are parasitic resistances of NMOS and PMOS transistors. During the charging phase ($t_{low}$) the input of the inverter is low and current passes through the PMOS and the output resistor. During the discharging phase ($t_{high}$), the input of the inverter is high and current goes through the output resistor and NMOS. As a result, the resistance values can be calculated from the lengths of time of each phase:
\begin{equation}
	R_{n}^{*}+R_{out}^{*} = (R_{n}+R_{out}) \cdot \frac{t_{low}+t_{high}}{t_{high}};
\end{equation}
\begin{equation}
	R_{p}^{*}+R_{out}^{*} = (R_{p}+R_{out}) \cdot \frac{t_{low}+t_{high}}{t_{low}}.
\end{equation}

Assuming that $R_{n}\approx R_{p}$ (this transistor balancing can be achieved by the appropriate relative sizing of the PMOS and NMOS transistors, for instance, by setting the PMOS width to 2.7 times the NMOS width for the UMC65nm technology) and $GND=0$, the equation~\ref{eq:voltage_divider} is simplified to:
\begin{equation}
\label{eq:voltage_divider_dc}
	V_{out} = V_{dd} \cdot \frac{t_{low}}{t_{low}+t_{high}} = V_{dd} \cdot (1 - DC),
\end{equation}
where $DC$ is the input duty cycle - the ratio between the length of time when the input clock is high during a clock period and the length of the clock period. 

Figure~\ref{fig:inv_vout_dc} shows the relationship between the input duty cycle and the output voltage of the PWM inverter. In the case when there is no output resistor, the dependency of the output voltage on the input is not linear. The reason of this non-linearity is that the PMOS and NMOS resistances change with the change of their drain voltages. Thus, $R_{p} \neq R_{n}$ when the value of $V_{out}$ is different from $V_{dd}/2$. This non-linearity, given the arithmetic functional requirements of a perceptron, is undesirable and needs to be either removed or compensated for. Compensation means very high per-inverter overheads which need to be precise in the analogue domain. However, by adding an output resistor $R_{out} \gg (R_{p}, R_{n})$, the difference between PMOS and NMOS resistances no longer affects the output, and the input duty cycle to output voltage relationship becomes completely linear. This requires no high-precision tuning in the analogue domain. 

\subsection {PWM Arithmetic}
\label{sec:pwm_arithm}

A perceptron needs to perform arithmetic operations. Converting from PWM to voltage is not the only function of the PWM inverters. They can also be used to construct arithmetic units, such as adders and weighted accumulators. Below we discuss these two operations and their circuits relevant to NNs. 

\begin{figure}[ht]
    \centering
    \includegraphics[width=0.7\textwidth]{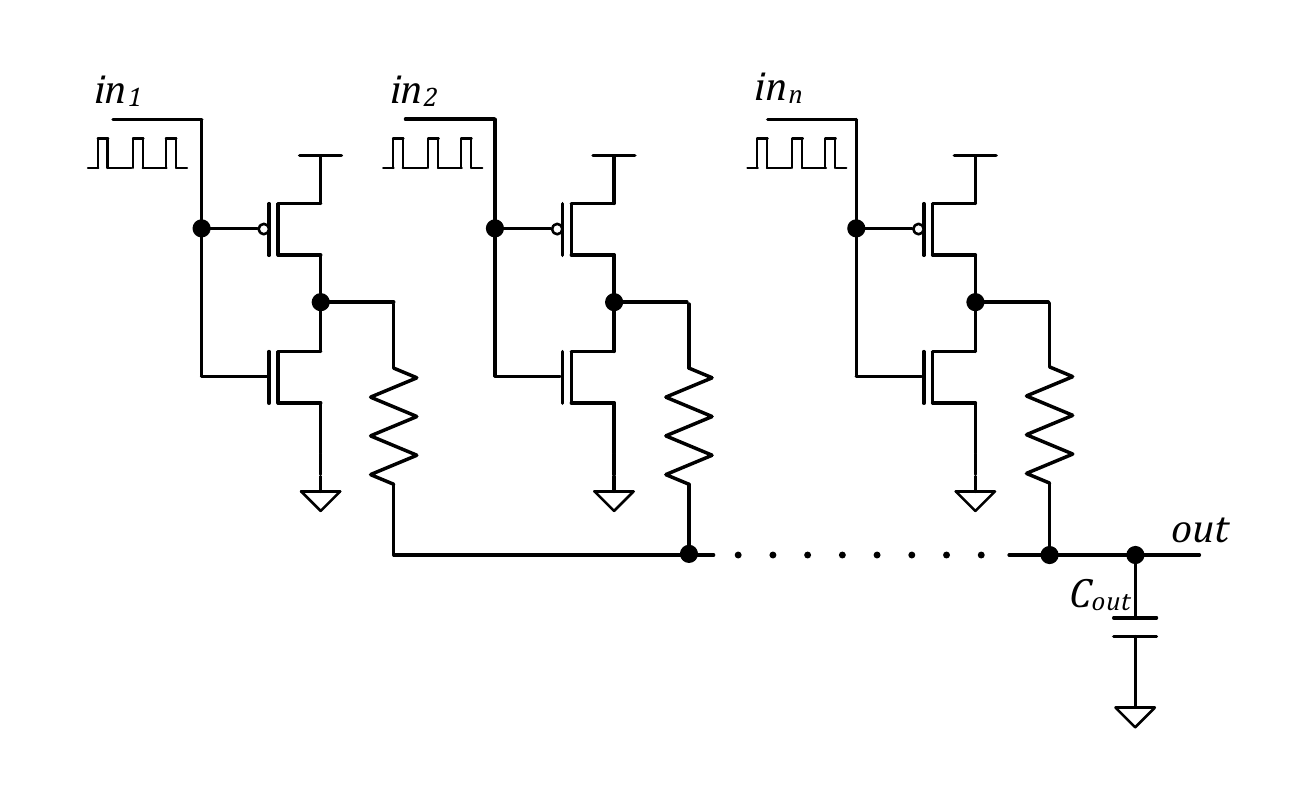}
    \caption{PWM adder circuit performed by parallel inverters, with outputs connected via a capacitor.}
    \label{fig:inv_sum}
\end{figure}

\paragraph{PWM Adder} The circuit of a PWM adder is shown in Figure~\ref{fig:inv_sum}. To add $n$ PWM-coded numbers we use $n$ inverters connected in parallel. Each inverter has an output resistor. The result is stored in the output capacitor in the form of its average voltage.

This kind of adder works on the principle of current summation and charge (i.e. voltage) accumulation. In other words, the values encoded in the input PWM signals are accumulated in the voltage on the output capacitor, and such circuits can be called voltage accumulators (VACs). To calculate the VAC output voltage, we use the principle of current summation and rewrite~\ref{eq:voltage_divider} using conductances instead of resistances. The following equation is for a single PWM inverter:
\begin{equation}
	V_{out} = V_{dd} \cdot \frac{G_{p}^{*}}{G_{p}^{*}+G_{n}^{*}},
\end{equation}
where $G_{p}^{*}=\frac{1}{R_{p}^{*}+R_{out}^{*}}$ and $G_{n}^{*}=\frac{1}{R_{n}^{*}+R_{out}^{*}}$.

Likewise equation~\ref{eq:voltage_divider_dc} can be expressed as follows:
\begin{equation}
    \label{eq:Gp}
	G_{p}^{*}= G \cdot \frac{t_{low}}{t_{low}+t_{high}} = G \cdot (1 - DC),
\end{equation}
\begin{equation}
    \label{eq:Gn}
	G_{n}^{*}= G \cdot \frac{t_{high}}{t_{low}+t_{high}} = G \cdot DC,
\end{equation}
where $G = \frac{1}{R_{p}+R_{out}} = \frac{1}{R_{out}+R_{n}}$.

Since the inverters in Figure~\ref{fig:inv_sum} are connected in parallel, the output voltage of a multi-inverter VAC can be given by:
\begin{equation}
\label{eq:voutGpGn}
	V_{out} = V_{dd} \cdot \frac{\sum_{i=1}^{n} G_{pi}^{*}}{\sum_{i=1}^{n} (G_{pi}^{*}+G_{ni}^{*})}.
\end{equation}

Using equations~\ref{eq:Gp} and~\ref{eq:Gn}, equation~\ref{eq:voutGpGn} can be simplified as:
\begin{equation}
    \label{eq:V_sum}
	V_{out} = V_{dd} \cdot (1 - \frac{\sum_{i=1}^{n} DC_{i}}{n}).
\end{equation}

In simple terms, the output voltage of a multi-inverter VAC is inversely proportional to the average value of the duty cycles of its inputs, which is exactly what is required.

\begin{figure}[ht]
    \centering
    \includegraphics[width=0.7\textwidth]{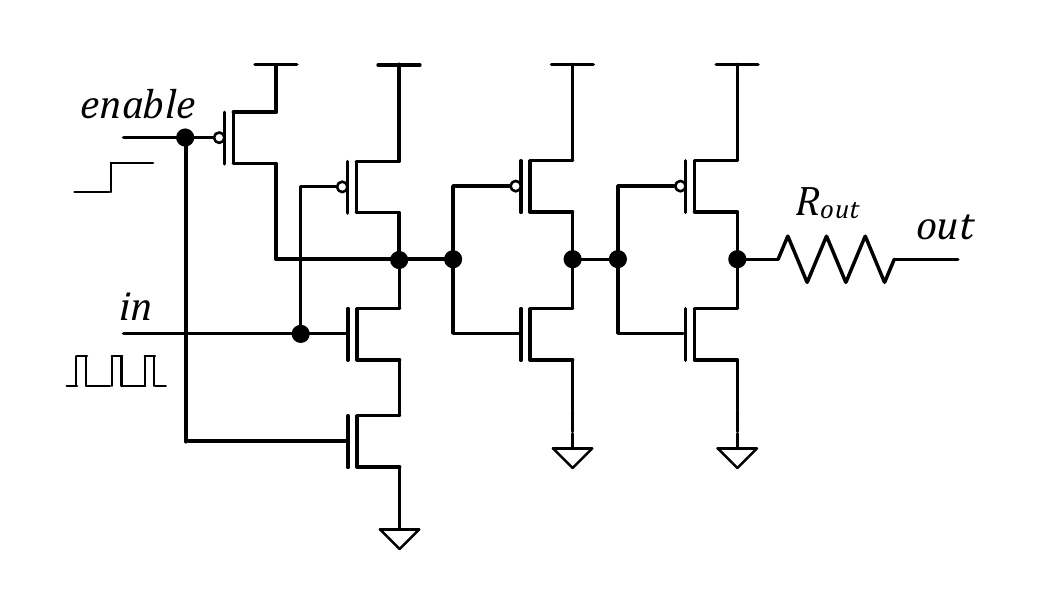}
    \caption{A single cell of the PWM weighted adder, based on a NAND gate.}
    \label{fig:and_res}
\end{figure}

\paragraph{Weighted PWM Accumulation} In order to design a perceptron, the ability to integrate weighted additions is another crucial design requirement. The VACs must be capable of programming the input weights, when required. This is performed by replacing the inverters by two-input NAND gates~(Figure~\ref{fig:and_res}). One input of this gate is  the PWM-coded signal, and the other is a digital switch signal for enabling or disabling this cell. The output of a disabled cell is always connected to $V_{dd}$ having the same effect as an enabled cell with zero input duty cycle. In this way, the perceptron can be programmed to determine which NAND gates participate in the accumulation. This programming may be carried out in the digital domain without affecting the voltage and frequency elasticity of the computation. 

\begin{figure}[ht]

    \centering
    \includegraphics[width=0.5\textwidth]{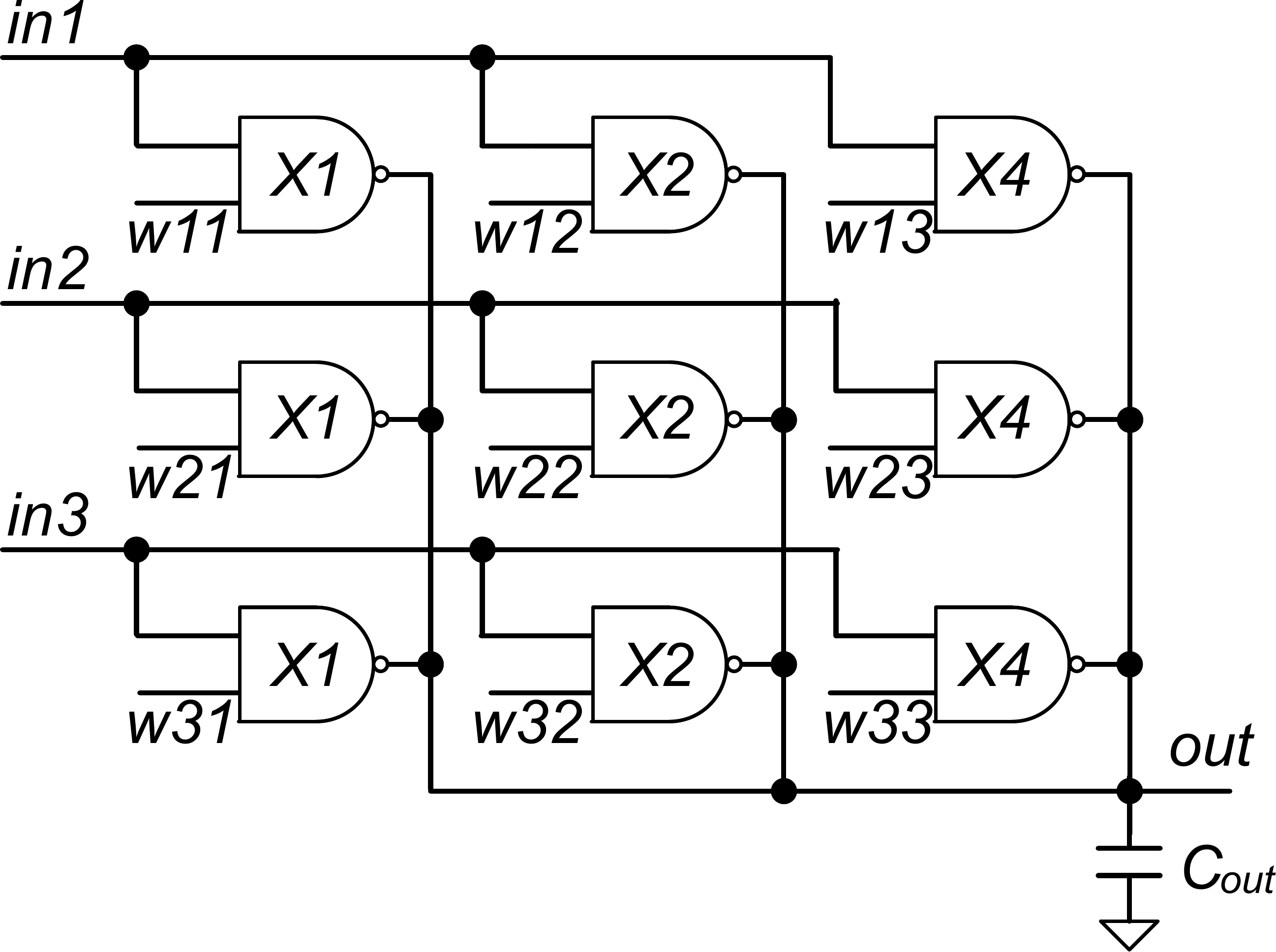}
    \caption{PWM weighted addition VAC with 3 inputs and 3-bit weights.}
    \label{fig:3x3}

\end{figure}

Figure~\ref{fig:3x3} shows a perceptron arithmetic VAC architecture for $3\times 3$ weighted addition based on these types of gates. As can be seen, the circuit adds 3 PWM-coded inputs multiplied by 3-bit weights. Every weight bit is implemented on a separate cell. The least significant bit goes to the cells with the smallest transistor sizes and the largest output resistors~(cells '$\times 1$'). The second bit is computed at the cells with doubled transistor widths and halved output resistances~(cells '$\times 2$'). And the most significant bit is coded with $4$ times the transistor widths, and $1/4$ times the output resistances~(cells '$\times 4$').

The output voltage of the $3\times 3$ weighted addition VAC can be calculated using~\ref{eq:V_sum}, considering the $\times 2$ and $\times 4$ cells as $2$ and $4$ single cells respectively.

\begin{equation}
\label{eq:perc_output_formula}
	V_{out} = V_{dd} \cdot (1 - \frac{\sum_{i=1}^{n} DC_{i} \cdot W_{i}}{n \cdot (2^{k}-1)}).
\end{equation}
where $n$ is the number of inputs, $k$ is the number of bits of the weight, $DC_i$ is the duty cycle of the input~$i$, and  $W_i$ is the weight of the input~$i$.

In the case of the $3\times 3$ weighted addition VAC, where $n=3$ and $k=3$, the output voltage is:

\begin{equation}
\label{eq:3x3_output_formula}
	V_{out} = V_{dd} \cdot (1 - \frac{\sum_{i=1}^{3} DC_{i} \cdot W_{i}}{21}).
\end{equation}

The arithmetic part of this equation is the weighted sum of duty cycles $DC_{sum}$:

\begin{equation}
\label{eq:dc_sum}
	DC_{sum} = \frac{\sum_{i=1}^{3} DC_{i} \cdot W_{i}}{21}.
\end{equation}

Thus, the definition of the $3\times 3$ weighted addition VAC is that its output voltage is proportional to the weighted sum of its input duty cycles, which is exactly as required:

\begin{equation}
\label{eq:3x3_vac_vout}
	V_{out} = V_{dd} \cdot (1 - DC_{sum}).
\end{equation}

\subsection {Voltage to PWM Conversion}
\label{subsec:VoltageToPWM}

In order to design a perceptron based on the type of VAC described in Section~\ref{sec:circuit_design}\ref{sec:pwm_arithm}, we need to provide an output interface for it. The output of the perceptron must be used as an input for the perceptrons of any subsequent layer in an NN. Therefore the output voltage of the PWM arithmetic unit (its VAC) should be converted back to the PWM format.

\begin{figure}[ht]
    \centering
    \includegraphics[width=0.6\textwidth]{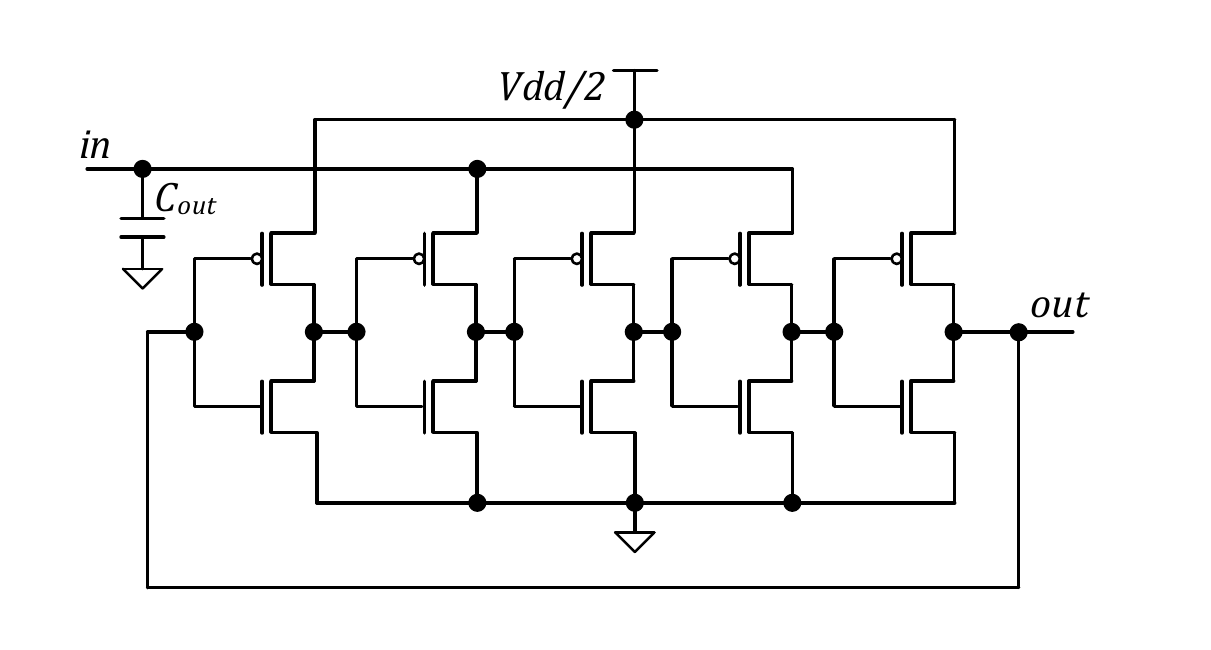}
    \caption{The ring oscillator based voltage to PWM converter.}
    \label{fig:v_to_pwm}
\end{figure}

The schematics of the voltage to PWM converter is shown in Figure~\ref{fig:v_to_pwm}. The converter circuit was proposed originally by~\cite{6705642}. The converter is a ring oscillator with different power supplies: the odd-numbered inverters are supplied with a voltage of $V_{dd}/2$, and the even-numbered inverters are supplied with the input voltage, which is the output voltage of the VAC. The difference between the supply voltages of the odd- and even-numbered inverters determines the output duty cycle. If the input voltage equals $V_{dd}/2$, the inverters have equal delay and the output duty cycle is $50\%$. If the input voltage increases, the period of switching from $0$ to $1$ increases, and the output duty cycle goes down. If the input voltage is lower than $V_{dd}/2$, the switching from $1$ to $0$ takes more time, and the output duty cycle goes up.

Given that the VAC theoretically achieves a linear relationship between its input duty cycle and its output analogue average voltage, the voltage to PWM converter should also ideally achieve a linear conversion relationship. In that case, the overall relationship between the input duty cycle signal and the output duty cycle signal would also be linear, for the simple case where the perceptron is programmed to do no arithmetic processing. In theory, the inverter chain-based voltage to PWM converter should be able to achieve this if the inverters are set to work in the linear regions of their transistors. 

\subsection {PWM-coded Perceptron Design}
\label{sec.perceptron_design}
The PWM-based perceptron consists of two main parts. The first part is the PWM arithmetic unit in the form of a VAC. This converts the PWM-coded inputs to a voltage which encodes the result of the computation as programmed by the enable signals. The second part then converts this voltage result to PWM format for use as inputs by subsequent perceptrons as their inputs. 
\begin{figure}[ht]
    \centering
        \centering
        \includegraphics[width=0.5\textwidth]{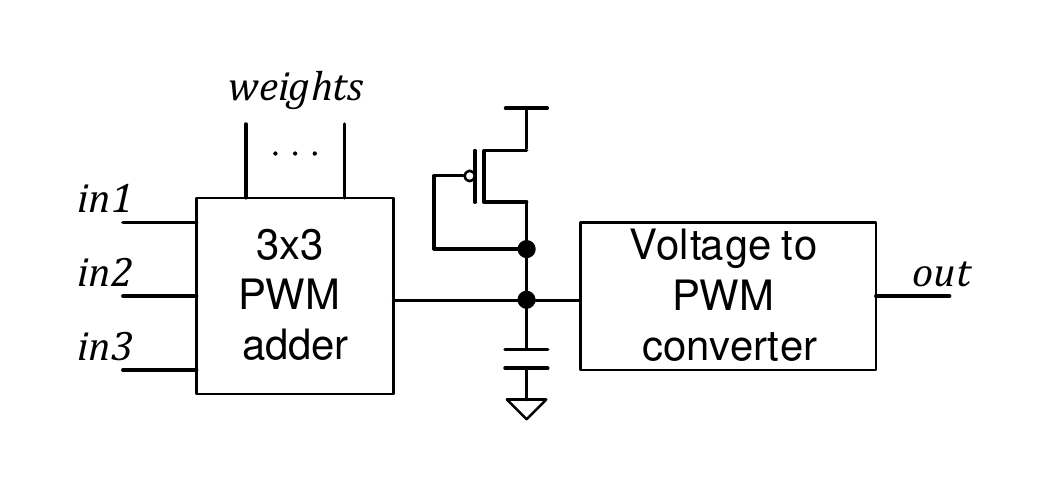}
        \caption{Structure of the perceptron: PWM adder, voltage to PWM converter, and compensation transistor.}
        \label{fig:perc_str_full}
\end{figure}        

This structure is shown in Figure~\ref{fig:perc_str_full}, with the $3\times 3$ weighted addition VAC as an example PWM-based arithmetic unit. Any desired VAC arithmetic unit can be put in this place to satisfy specific perceptron functionality requirements. The simple glue logic consisting of a PMOS transistor between the two blocks will be discussed in detail in Section~\ref{sec:results}\ref{subsec:perc_results}.

The size of such a perceptron is such that its design may be entirely analysed and validated through simulations within the VLSI CAD environment in which it is implemented. At least some of this analysis must be conducted in the analogue signal domain as the voltage signal between the two parts of the perceptron holds the computation results in its analogue value. As a result, simulations in a VLSI CAD tool environment that support mixed-signal or analogue studies are the best way of analysing and validating such designs. In this work we implement our perceptron and analyse it using the Cadence Analogue Design Environment. Detailed results will be shown in Section~\ref{sec:results}\ref{subsec:perc_results}.

\subsection {PWM-coded Neural Network Design}
\label{subsec:pwm_based_nn}

\begin{figure}[ht]
    \centering
    \begin{minipage}[b]{1\linewidth}
        \centering
        \includegraphics[width=\textwidth]{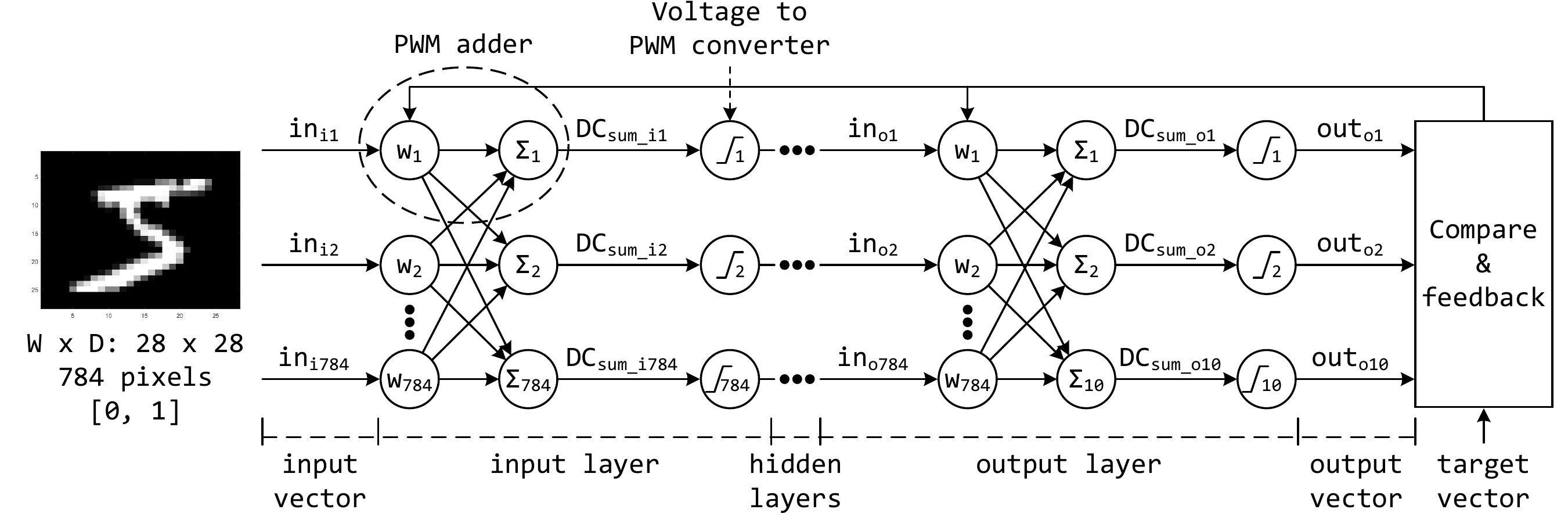}
        \caption{Neural network for MNIST. The $DC_{sum}$ signals are voltages. $in$ and $out$ signals are duty cycles.}
        \label{fig:nn}
    \end{minipage}
    \hspace{0.5cm}
\end{figure}

The proposed PWM perceptrons can be used in constructing traditional NNs such as the example shown in Figure~\ref{fig:nn}. In this NN, the input vector ($in$) is fed to the input layer, and the activity propagates through a number of hidden layers to reach the output layer, where the output vector ($out$) is generated.
Then, the output vector is compared to the target vector and the error is back propagated to update the weights of each layer using gradient descent.
This procedure is iterated with respect to the specified epoch.

In this work, the $in$ and $out$ signals are of the PWM-type. The value of such a signal, which is between 0 and 1, is represented by its duty cycle value between 0\% and 100\%.
The VAC arithmetic units then compute on such $in$ values. This is illustrated by the example described by equation~\ref{eq:dc_sum}, where each $in$ is multiplied by its weight and all results are accumulated by the VAC in the $DC_{sum}$ voltages. In other words, the weight and sum blocks in Figure~\ref{fig:nn} are implemented by the proposed perceptron's VAC. 
Then, every $DC_{sum}$, which is an analogue voltage across a capacitor, is fed to the activation function (AF) whose output is in PWM format to be used as the input of the next layer. 
To include the AF, equation~\ref{eq:dc_sum} can be modified as expressed in equation~\ref{eq:mac}. This requires that the voltage to PWM conversion also implements the AF. Potential modifications from the basic ring oscillator may be necessary, although the basic ring oscillator already approximates a popular AF. This will be discussed in detail in Section~\ref{sec:results}\ref{subsec:perc_results}.
 
Finally, $out$ is obtained, the error is calculated and every weight is adjusted by the back-propagation (BP) algorithm. The comparison to target vector is not necessarily implemented with a perceptron-like device and may be implemented by some external controller, which is outside the scope of this paper.

For the PWM-coded NNs, a number of design choices must be made: weight encoding, maximum weight, AF and number of layers, among others.
This section establishes a method of making the best use of the proposed PWM-based perceptron to construct NNs to perform specific computational tasks. 
We will explore aspects of NN design, including weight types, AF, maximum weight and number of layers. 
We will use the well-known handwriting digit (MNIST) classification problem~\cite{726791}, which is widely used for machine learning algorithm testing~\cite{Du-2018-TCAD}, as the benchmark application and case study for this investigation. 
The goal is to suitably determine the best NN configurations for the proposed PWM-coded NN. 

\begin{equation}
\label{eq:mac}
    out = f(DC_{sum}) = f(\frac{\sum_{i=1}^{n} in_{i} \cdot W_{i}}{n \cdot (2^{k}-1)})
\end{equation}

\paragraph{Integer Weight and Training}
\label{subsubsec:interger_weight}
Regarding the circuit design, the weight is discretised to an integer value. This is different from most related work where floating-point (FP) numbers are used for weights.
As the circuit size depends on the bit-width of the weight, it is crucial to find the smallest bit-width that still provides the specified error rate tolerance.

\begin{figure}[ht]
    \centering
    \begin{minipage}[b]{0.47\linewidth}
        \centering
        \includegraphics[width=\textwidth]{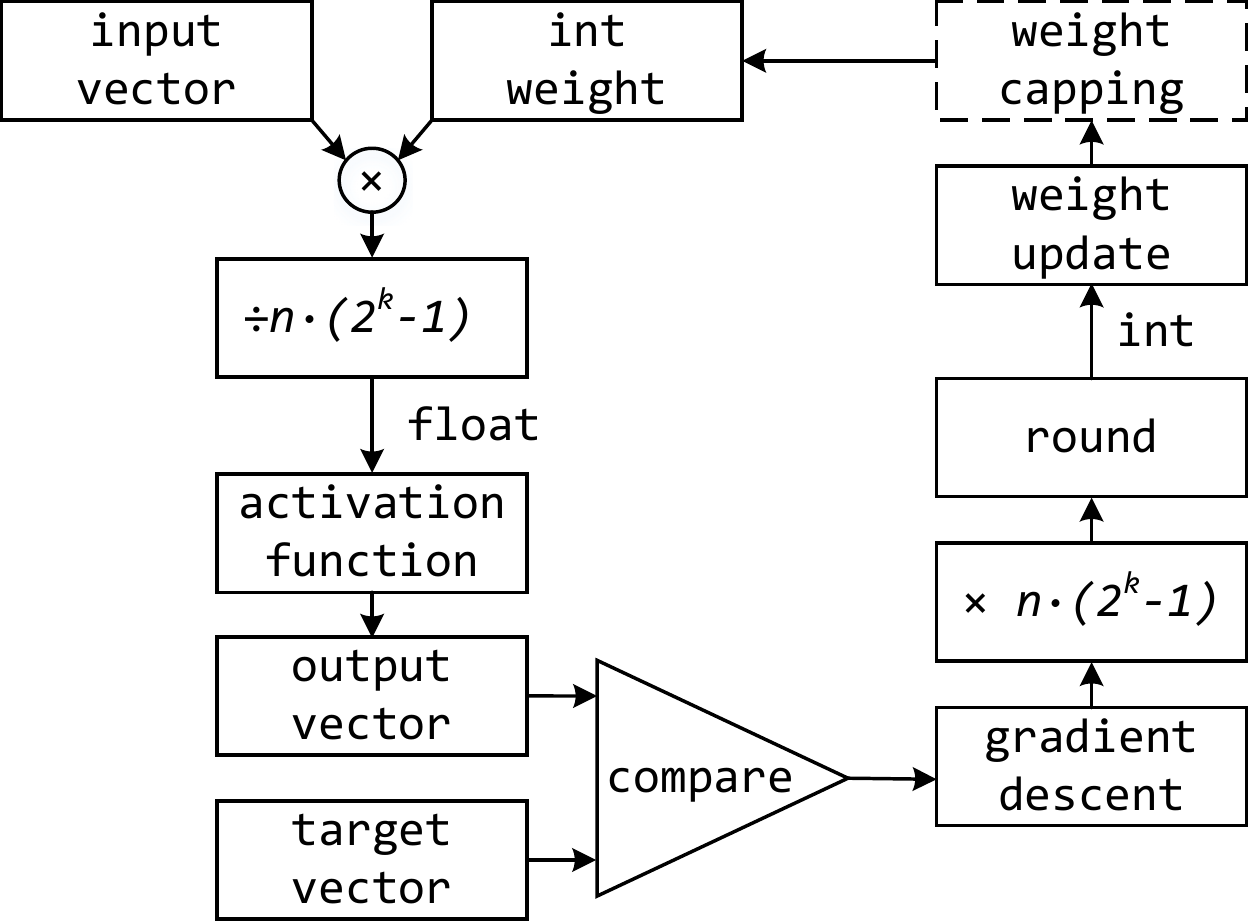}
        \caption{Integer weight training.}
        \label{fig:nn_training}
    \end{minipage}
    \hspace{0.5cm}
\end{figure}

The integer weight training can be designed as illustrated in Figure~\ref{fig:nn_training}.
The MNIST input vector ($in$) is multiplied by the integer weight ($W$) and the results are accumulated as $out$.
Then, $out$ is divided by $n \cdot (2^{k}-1)$ (i.e. normalising), which yields the final value of $out$ between 0 and 1.
Consequently, $out$ is scaled to the same range and comparable to the target vector. 
Then, $out$ passes the AF, and the output vector is obtained and compared to the target vector.
Next, the FP update is computed from the gradient descent, the learning rate, and the error.
To adjust the integer weight, the update is scaled back to the integer number by multiplying by $n \cdot (2^{k}-1)$ and rounding.
Next, the integer weight is updated and capped if it exceeds the specified bit-width (e.g. the example $3\times 3$ weighted addition VAC in Figure~\ref{fig:3x3} has 3-bit weights).
Finally, the training process iterates until the number of specified epochs is reached.
Note that the weight capping can be disabled to allow  unlimited weight adjustments to mimic FP training.

\paragraph{Activation Function}
\label{subsubsec:act_fn}
The AF is necessary in an NN-based learning process because it provides non-linearity to the computation so that the learning is not limited to  linear problems. It also helps map the resulting values in a certain range, depending on the function.

In this work, the input and output ranges of the AF are a main concern because they need to match the output format of the problem and the circuit behaviour. In other words, depending on the purpose of the NN, it may expect its input and output variables to take values within certain ranges. These ranges then need to be mapped onto the working signal range of our perceptron, which is restricted by the duty-cycle representation between 0\% and 100\%. Here we take popular MNIST benchmark~\cite{deng2012mnist} as an exemplar to explore this aspect of NN design using our perceptron as the basic building block.

In the context of MNIST, the AFs are needed to provide a fully positive output to comply with the target vector~\cite{726791}.
Also, our perceptron design stores the VAC result as the voltage across $C_{out}$ between its two blocks, which means that the $DC_{sum}$ signals are entirely positive voltages. And such a voltage gets converted to a PWM duty cycle, which is also entirely positive. 
For these reasons, the well-known AF ReLU~\cite{Hagan:1997:NND:249049,6638312}, which has an entirely positive output range, is best suited. 

Certain other popular AFs are less suitable for this initial investigation. For instance, the sigmoid function is clearly non-linear across an input range between -5 and 5~\cite{Hagan:1997:NND:249049}, which requires representation of negative values. The non-linearity also means that major modifications to the voltage to PWM part need to be investigated for implementing such AFs. hence, we decided to concentrate on trying to mimic the ReLU AF using our perceptron's voltage to PWM converter. 

\begin{figure}[ht]
    \centering
    \begin{minipage}[b]{0.45\linewidth}
        \centering
        \includegraphics[width=\textwidth]{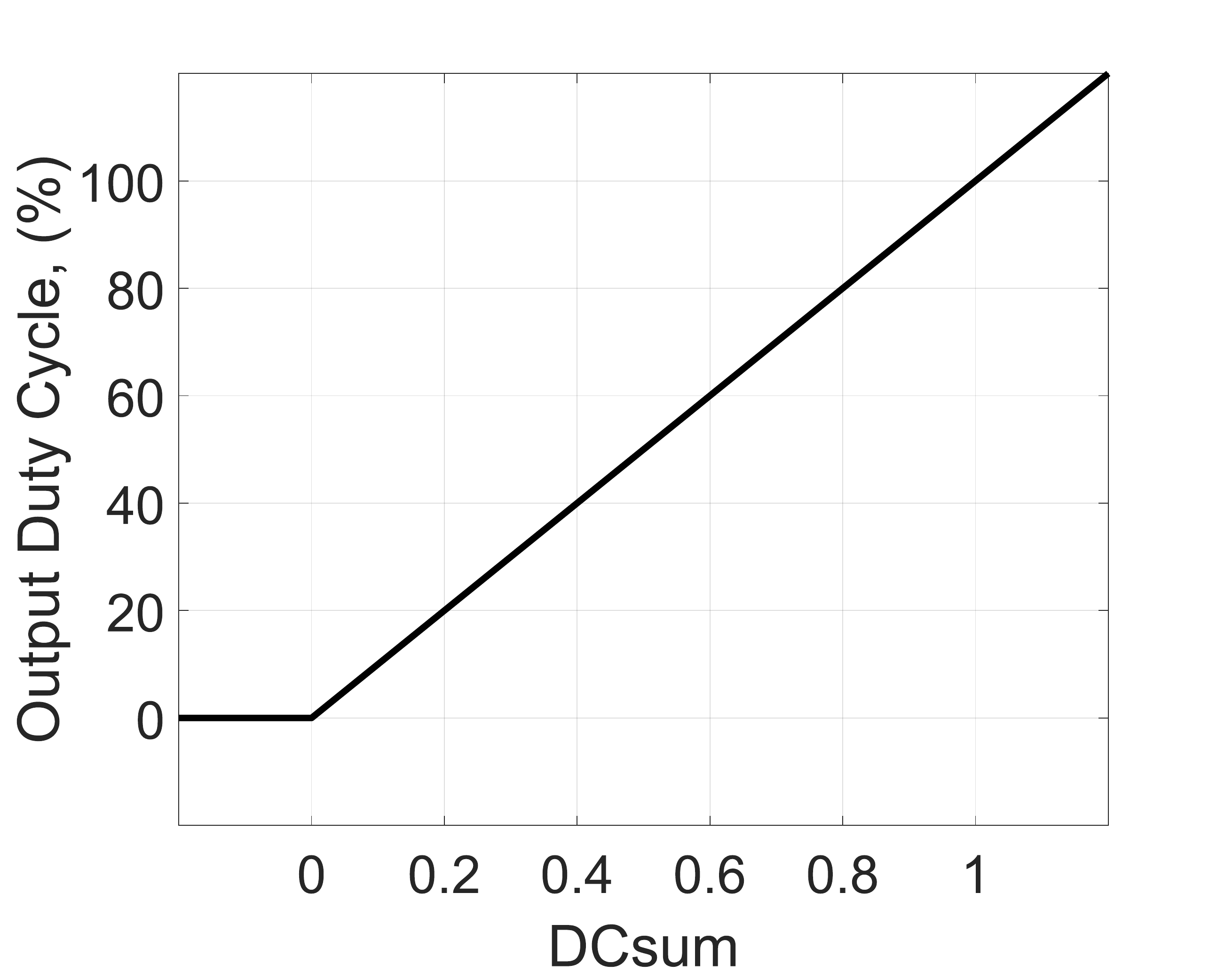}
        \caption{ReLU function.}
        \label{fig:relu}
    \end{minipage}
    \hspace{0.5cm}
    \begin{minipage}[b]{0.45\linewidth}
        \centering
        \includegraphics[width=\textwidth]{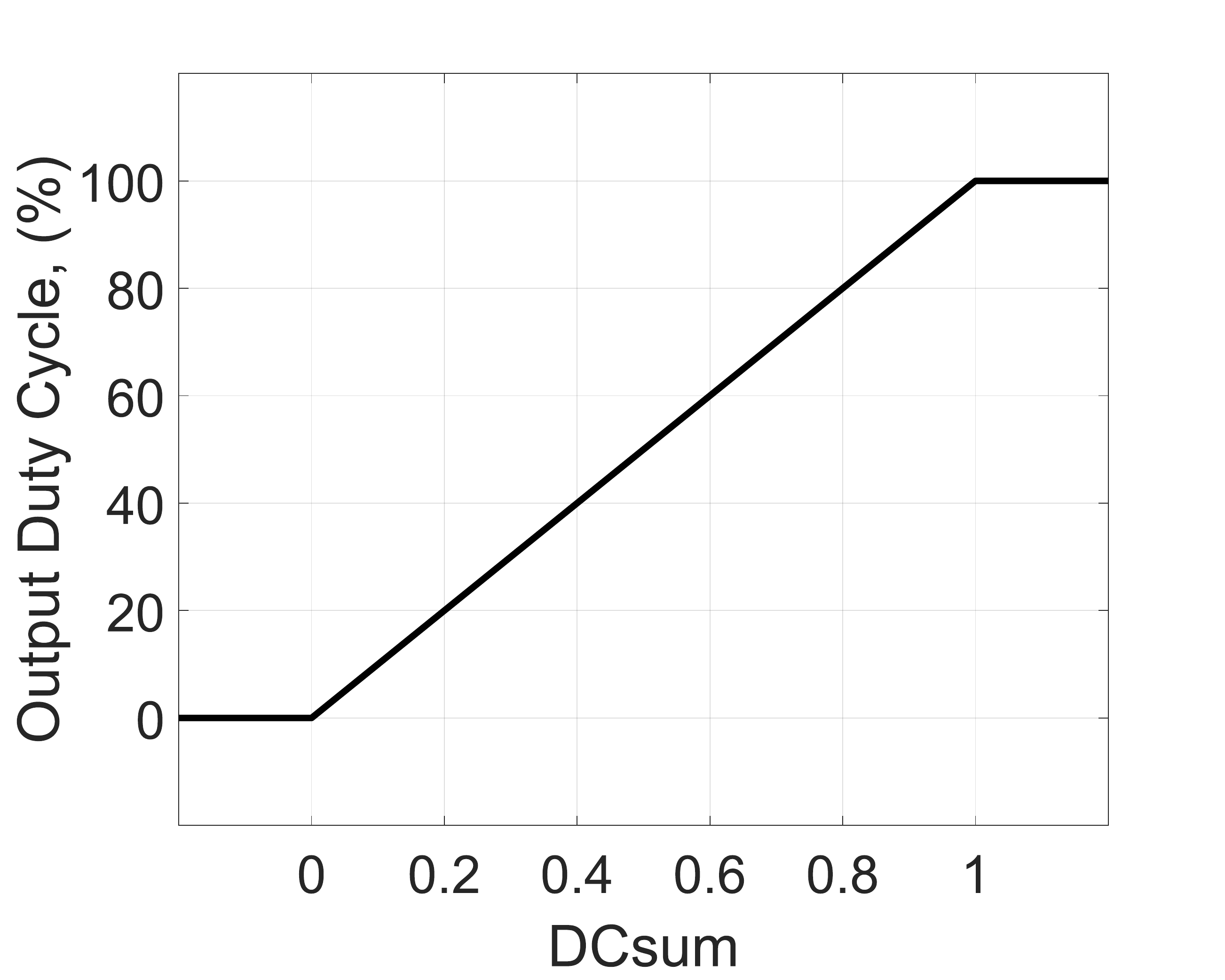}
        \caption{Capped ReLU function.}
        \label{fig:relu_capped}
    \end{minipage}
\end{figure}

The ReLU function in equation~\ref{eq:relu}~\cite{6638312} is depicted in Figure~\ref{fig:relu}.
One of its attractions is that it is easily differentiable, facilitating gradient descent.
To mimic the output of the VAC, it is better than the sigmoid function because the charge in the output capacitor ($C_{out}$) is emptied when the VAC result is negative. 
Otherwise, the capacitor is charged and the positive result is obtained.
However, the output of this function must be capped at 1 to represent the limit of the PWM range as shown in equation~\ref{eq:cap_relu} and  Figure~\ref{fig:relu_capped}.
This work will attempt to construct an AF that approximates the capped ReLU function. 

The size of an entire NN designed for the MNIST problem is such that it is not possible to analyse it entirely within a VLSI CAD environment. For instance, to analyse an image of 784 pixels (cf. the example in Fig.~\ref{fig:nn}) there need to be 784 perceptrons in the first layer of the NN alone and this is clearly beyond analogue simulations at the VLSI level. Effort must be expended in building models in a higher-level language to investigate the design properly. 

\begin{equation}
\label{eq:relu}
f(x)=
\begin{cases}
0 &, x<0
\\
x &, x>0
\end{cases}
\end{equation}
\begin{equation}
\label{eq:cap_relu}
f(x)=
\begin{cases}
0 &, x<0
\\
x &, 0<x<1
\\
1 &, x>1
\end{cases}
\end{equation}

\section{Results}
\label{sec:results}
This section reports experimental results of the PWM-based perceptron, leading to an NN architecture. These results validate the design methods at both circuit- and architecture-level. 

\subsection {Analysis and Validation of PWM-coded Perceptron}
\label{subsec:perc_results}

A prototype circuit of the PWM perceptron is designed using UMC65nm technology and simulated in the Cadence Analog Design Environment tool\footnote{URL: https://tinyurl.com/y6k73k4t}. We used the high voltage transistors (with $2.5V$ nominal voltage) in purpose of better observation.
Below we analyse the behaviour of the perceptron circuit under different parametric variations, generated by the design tool.

\begin{figure}[ht]
    \centering
    \includegraphics[width=0.7\textwidth]{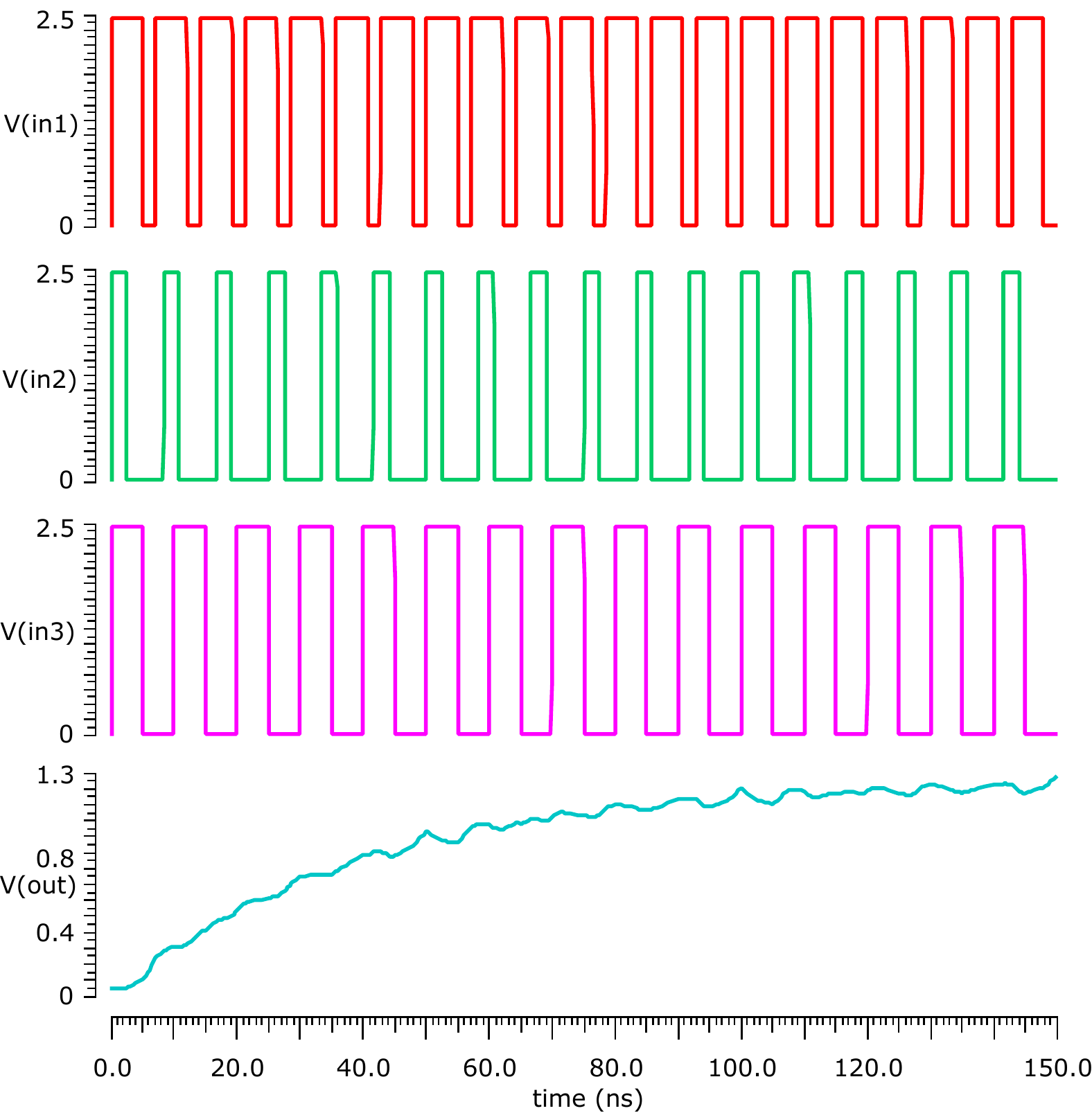}
    \caption{Capacitor charging in the 3 inverters VAC.}
    \label{fig:3_inv}
\end{figure}

\paragraph {1. VAC Validation}
\label{subsubsec:vac_results}
The first constituent part of the perceptron is the VAC. Figure~\ref{fig:3_inv} shows the charging of the capacitor in the VAC based on three inverters connected in parallel as shown if Figure~\ref{fig:inv_sum}. The frequencies and duty cycles of the inputs are: $f_1=140MHz$, $DC_1=70\%$, $f_1=120MHz$, $DC_1=30\%$, $f_1=100MHz$, $DC_1=50\%$. The capacitor have been charged to the voltage value, proportional to the average duty cycle of the inputs. The charging time of the capacitor depends on the $RC$ value, and the input frequency does not affect it. However, if the frequency is too low, it may result in too high ripple of the output voltage, and, thereafter, reduction of accuracy.

To support our VAC design based on inverters/NANDs and voltage summation on a capacitor, we implemented the $3 \times 3$ weighted addition VAC shown in Figure~\ref{fig:3x3} in Cadence and ran simulation experiments on it. The results of these simulations are compared to theoretical results obtained from~\ref{eq:3x3_output_formula} and compared in Table~\ref{tab:3x3_simulations}. The differences between the theoretical and simulation results do not exceed 10\%. These results validate the correctness of the PWM-based weighted addition VAC design.

\begin{table}[h]
    \centering
    \caption{Experimental and theoretical results of the $3 \times 3$ weighted adder.}
    \label{tab:3x3_simulations}
    \begin{tabular}{| c | c | c | c | c | c | c | c | }
        \hline
        \multirow {2}{*}{DC1}         & \multirow {2}{*}{W1} & \multirow {2}{*}{DC2} & \multirow {2}{*}{W2} & \multirow {2}{*}{DC3} & \multirow {2}{*}{W3} & $V_{out}$  & $V_{out}$ \\
         & & & & & & theoretical &  simulation \\
        \hline
         70\% & 7 & 80\% & 7 & 90\% & 7 & 0.50V & 0.51V \\
        \hline
        50\% & 1 & 50\% & 2 & 50\% & 4 & 2.08V & 2.11V \\
        \hline
        20\% & 5 & 60\% & 6 & 80\% & 7 & 1.29V & 1.33V \\
        \hline
        95\% & 7 & 90\% & 6 & 80\% & 6 & 0.50V & 0.45V \\
        \hline
        30\% & 1 & 40\% & 4 & 50\% & 2 & 2.16V & 2.21V \\
        \hline
        80\% & 7 & 20\% & 3 & 50\% & 4 & 1.54V & 1.61V \\
        \hline
    \end{tabular}
\end{table}

\begin{figure}[h]
    \centering
    \begin{minipage}[b]{0.45\linewidth}
        \centering
        \includegraphics[width=\textwidth]{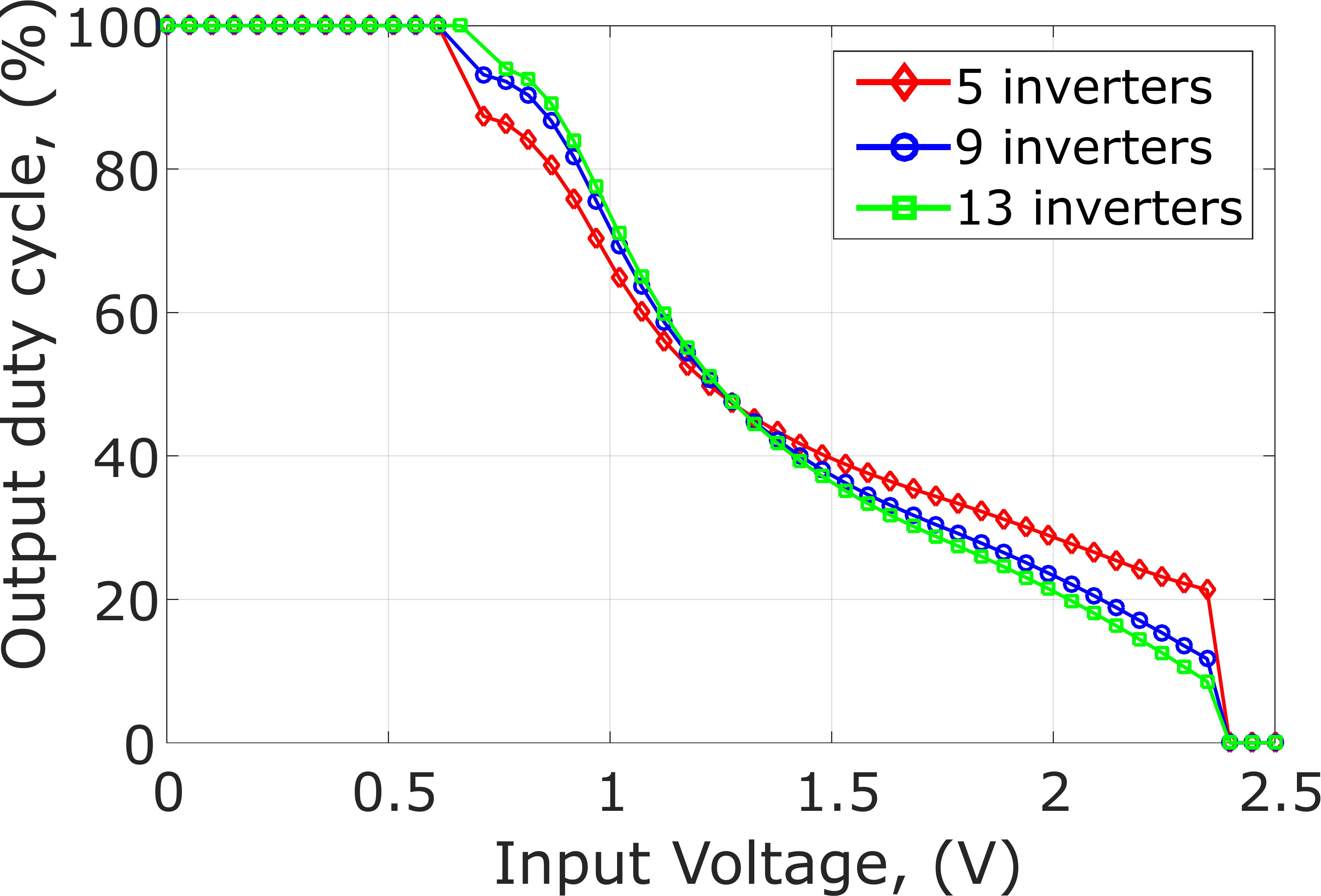}
        \caption{Output duty cycle of the voltage to PWM converter.}
        \label{fig:v_to_pwm_inout}
    \end{minipage}
    \hspace{0.5cm}
    \begin{minipage}[b]{0.45\linewidth}
        \centering
        \includegraphics[width=\textwidth]{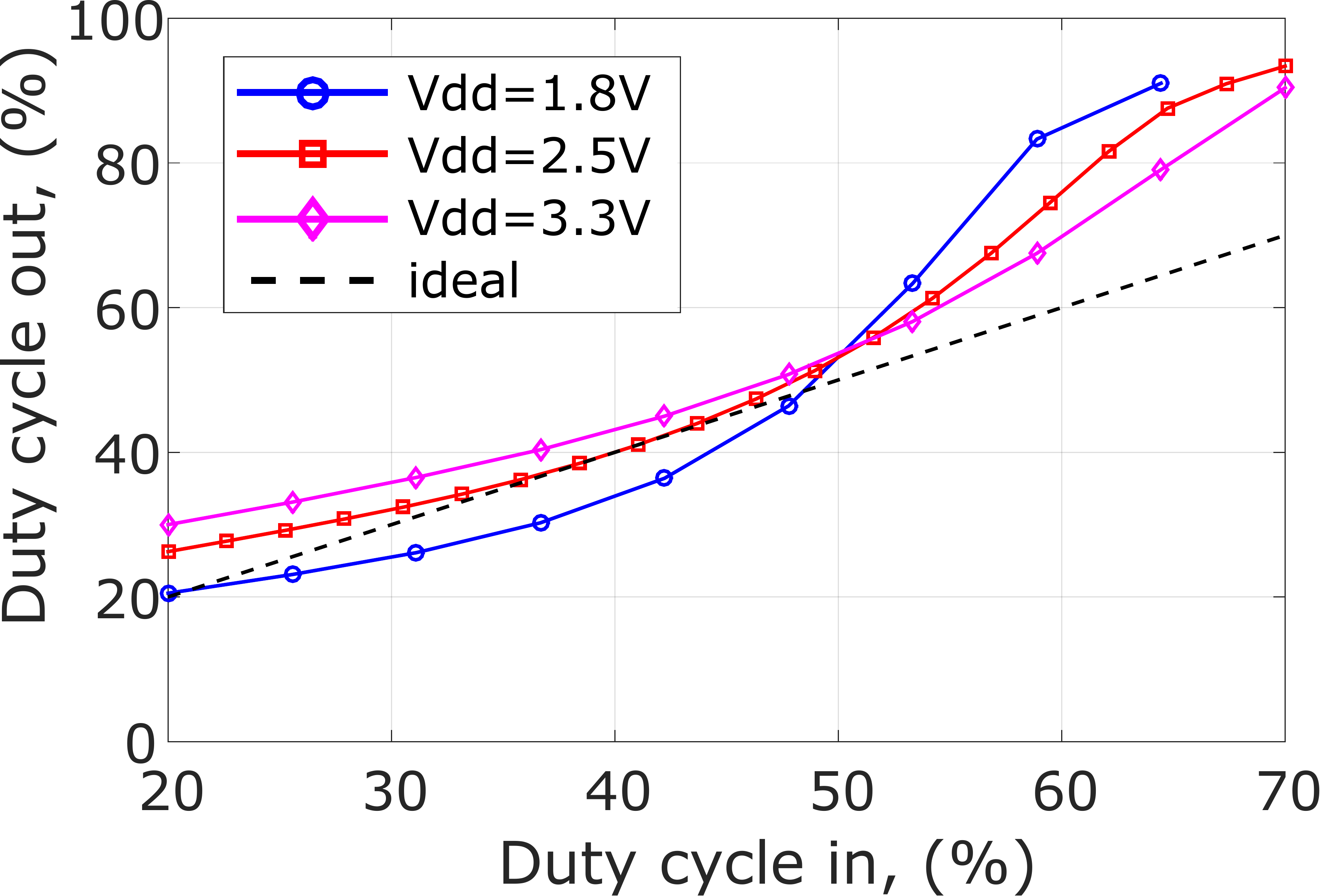}
        \caption{Output vs input duty cycle of the perceptron.}
        \label{fig:perc_inout}
    \end{minipage}
\end{figure}

\paragraph {2. Validation of Voltage to PWM Conversion}
\label{subsec:VoltageToPWM_results}
The second constituent part of the proposed perceptron is the voltage to PWM converter. This converts the result of VAC arithmetic computation stored as an analogue voltage (a $DC_{sum}$ signal) back to the PWM format for output to subsequent perceptrons, as presented in Section~\ref{sec:circuit_design}\ref{subsec:VoltageToPWM}. 
The Cadence Analog Environment simulation results of the voltage to PWM converter is shown in Figure~\ref{fig:v_to_pwm_inout}. Ideally, the voltage to PWM conversion should be linear. The real relationship between the output and the input is almost linear for input voltages between $0.7V$ and $2.3V$. However, outside this range the ring oscillator stops oscillating. The reason for this will be discussed below.

Another interesting effect is that the linearity of the output increases with increasing the number of inverters, but the difference between 9 and 13 inverters is small. Thus, a chain of 9 inverters should be considered a reasonable voltage to PWM converter and we use this design in our subsequent studies.

\paragraph {3. Perceptron Experimental Results and Design Adjustments}
\label{subsec:perceptron_results}

Figure~\ref{fig:perc_inout} shows the combined operation of both parts of the perceptron: the $3\times 3$ weighted PWM addition VAC (Figure~\ref{fig:3x3}) connected to the voltage to PWM converter (Figure~\ref{fig:v_to_pwm}). The three inputs of the perceptron are connected together, and all the weights are $7$ (all the cells are enabled). The line labelled 'ideal' is obtained through the equations in Section~\ref{sec:circuit_design}. Analysing these results we can say that:

\begin{itemize}

\item In this simulation for the ideal case we expect the output duty cycle to be equal to the input duty cycle. However, the real output is slightly different from the ideal; and this difference increases with the input duty cycle above $50\%$.

\item The output duty cycles for different supply voltages are similar. The difference does not exceed $10\%$. This indicates voltage variation resilience in the perceptron design.

\item The input duty cycle has limited range - from $20\%$ to $70\%$. Beyond this range the output stops oscillating and becomes a constant signal. 

\end{itemize}

The observed reduction of operational range and loss of linearity in the voltage to PWM converter are caused by the fact that the voltage $DC_{sum}$ powers the voltage to PWM converter. When the input duty cycle is above $70\%$, $DC_{sum}$ is below $30\%$ of $V_{dd}$. For $V{dd}=2.5V$ this is below the threshold voltage. And in this case the NMOS transistors of the ring oscillator are always off, and the output stops oscillating. In other words, there is a mismatch between the voltage ranges of the two parts of the perceptron. The output voltage range of the PWM weighted addition VAC is from $0V$ to $2.5V$ (Figure~\ref{fig:inv_vout_dc}); and the input voltage range of the voltage to PWM converter is from $0.7V$ to $2.3V$ (Figure~\ref{fig:v_to_pwm_inout}).

\begin{figure}
    \hspace{0.5cm}
        \centering
        \includegraphics[width=0.5\textwidth]{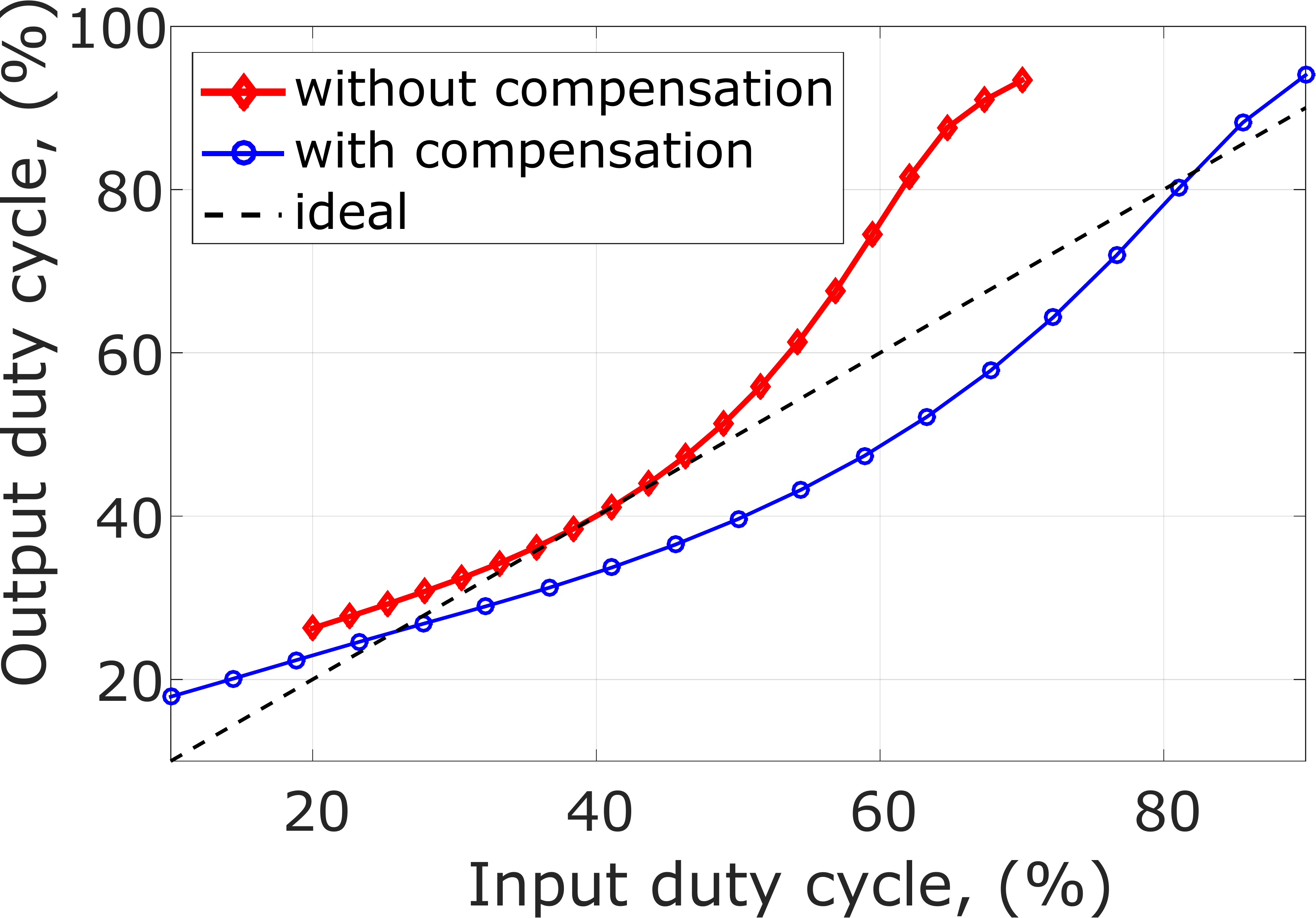}
        \caption{Output vs input duty cycle of the perceptron with and without compensation.}
        \label{fig:perc_comp}
\end{figure}

We may limit the range of the output voltage of the PWM weighted addition VAC. This can be done by adding a small glue logic between the two blocks of the perceptron. This may consist of no more than a compensation PMOS transistor, whose gate and drain are connected to the capacitor as shown in Figure~\ref{fig:perc_str_full}. In this case, when the voltage on the capacitor goes below the threshold, the PMOS starts charging this capacitor, and when the voltage is above the threshold, the PMOS is off.

The input and output duty cycles of the perceptron with compensation are depicted in Figure~\ref{fig:perc_comp}. The output is closer to the ideal, and its range is much wider: from $10\%$ to $90\%$.

\paragraph {4. Power Elasticity and Resilience}
\label{subsec:percPowerResillience_results}

\begin{figure}[h]
    \centering
    \begin{minipage}[b]{0.45\linewidth}
        \centering
        \includegraphics[width=\textwidth]{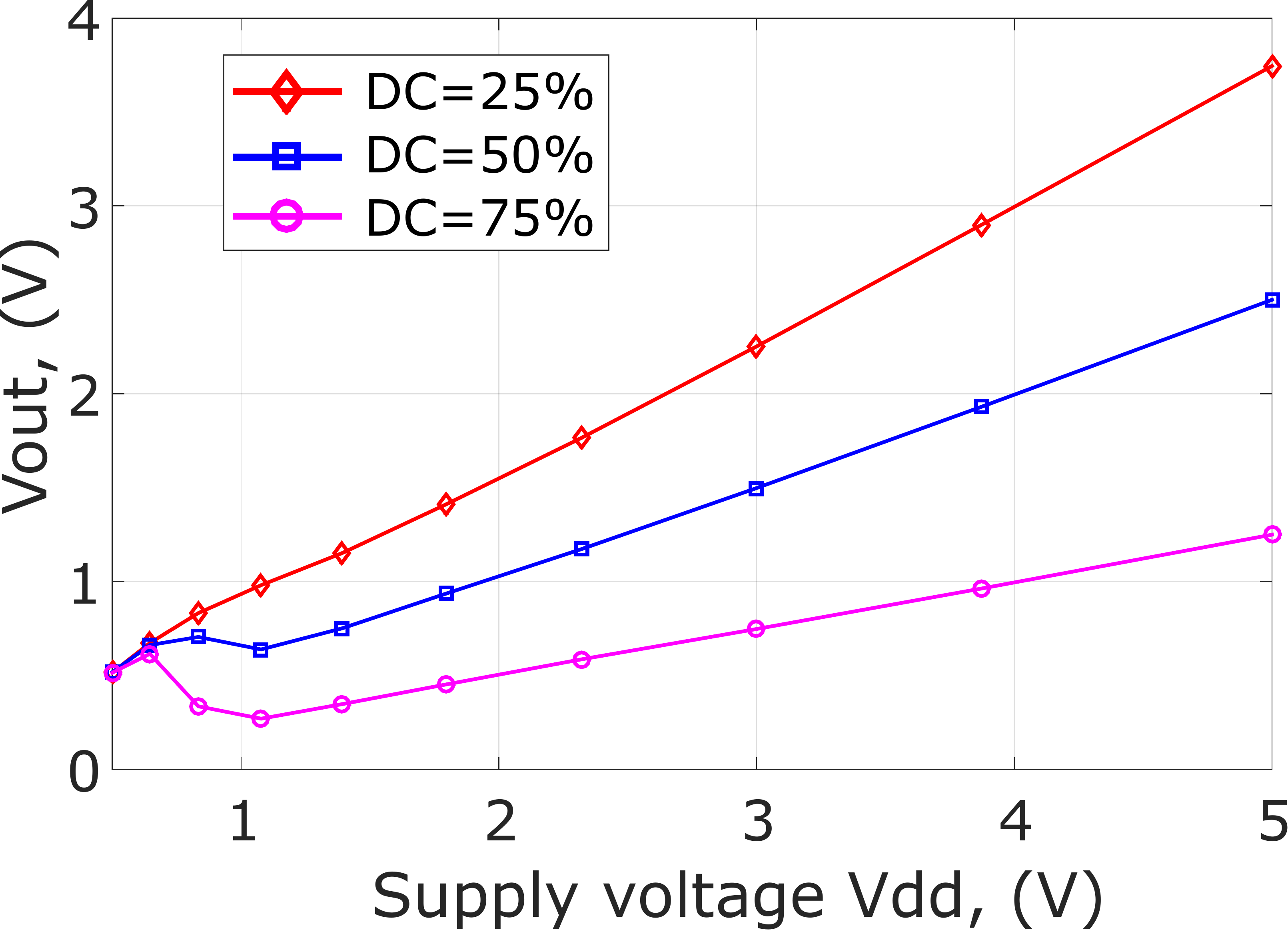}
        \caption{Output voltage (absolute values) vs static variation of power supply.}
        \label{fig:out_vs_vdd}
    \end{minipage}
    \hspace{0.5cm}
    \begin{minipage}[b]{0.45\linewidth}
        \centering
        \includegraphics[width=\textwidth]{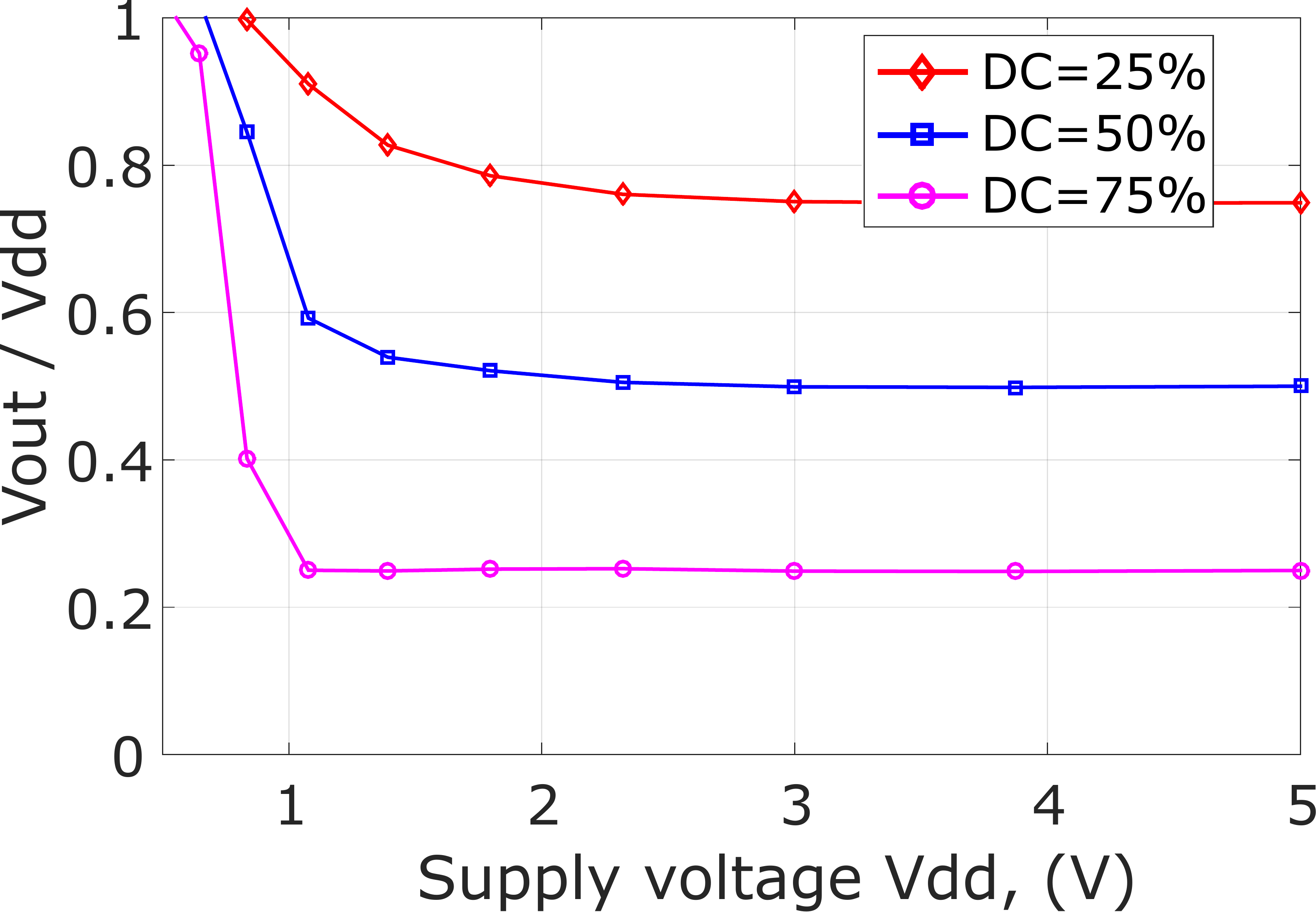}
        \caption{Output voltage (relative values) vs static variation of power supply.}
        \label{fig:out_vs_vdd_abs}
    \end{minipage}
\end{figure}

To demonstrate the perceptron's resilience to power variations we simulated the the $3\times 3$ PWM-based weighted addition VAC circuit (Figure~\ref{fig:3x3}) with different values of supply voltage and input signal amplitude. The results are shown in Figure~\ref{fig:out_vs_vdd}. As can be seen, the output voltage grows almost linearly with increased $V_{dd}$. As expected, higher duty cycle show lower output voltages, and vice versa. In the case of the unstable supply voltage, the absolute value of the output voltage does not bear any reliable information. In this case, we should consider the relative relationship between the output voltage and the supply voltage. This relationship should be proportional to the input duty cycle independently from $V_{dd}$. This is demonstrated by Figure~\ref{fig:out_vs_vdd_abs} where the $y$ axis represents not the absolute value of $V_{out}$, but the ratio between $V_{out}$ and $V_{dd}$ that is more relevant for unstable power conditions.

The circuit shows high resilience to static supply voltage variations. Starting from 1 - 1.5V the ratio $V_{out}$ and $V_{dd}$ remains the same for each duty cycle value of the input signal.

\begin{figure}[h]
    \centering
    \begin{minipage}[b]{0.45\linewidth}
        \centering
        \includegraphics[width=\textwidth]{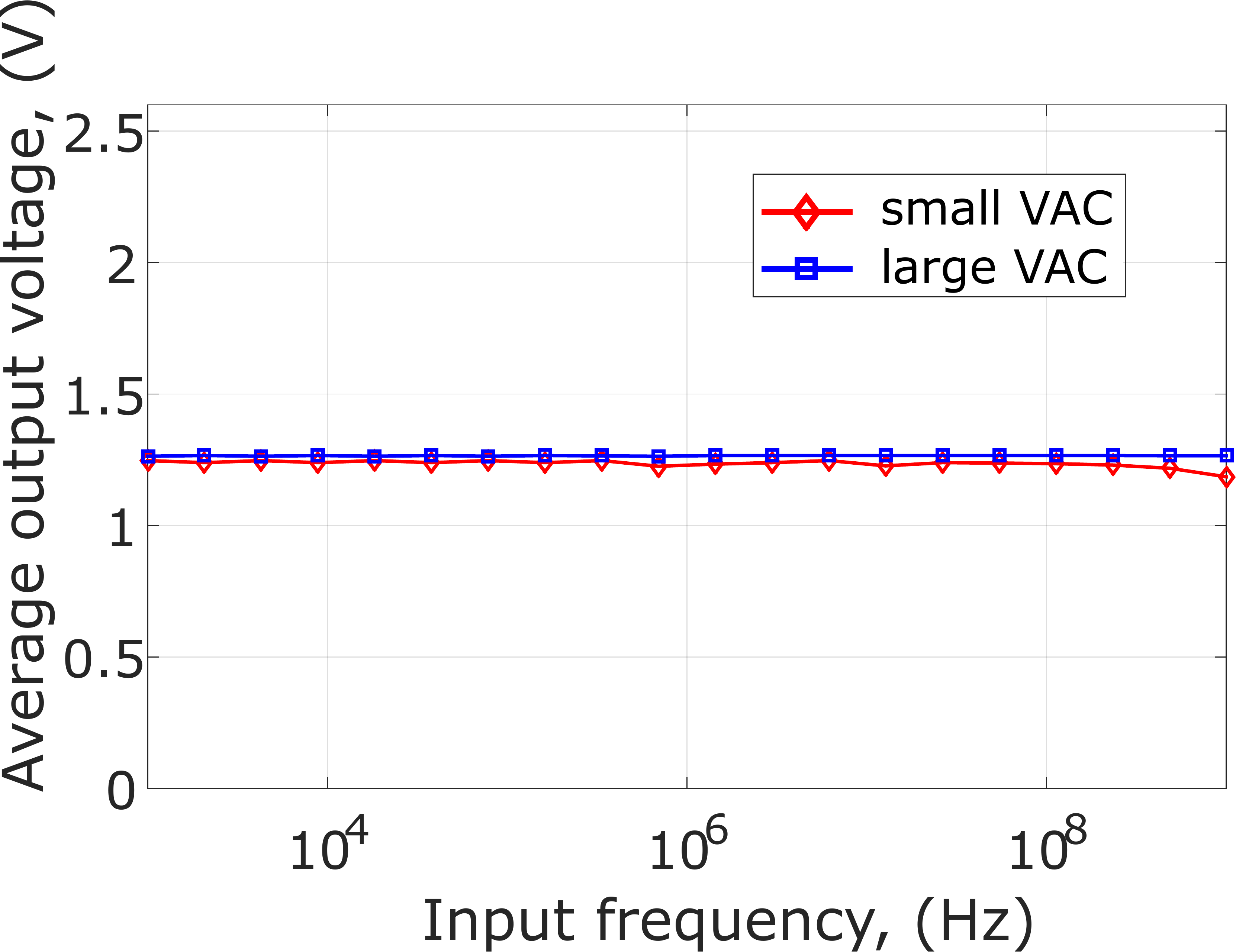}
        \caption{Output voltage vs static variation of input frequency.}
        \label{fig:v_vs_freq}
    \end{minipage}
    \hspace{0.5cm}
    \begin{minipage}[b]{0.45\linewidth}
        \centering
        \includegraphics[width=\textwidth]{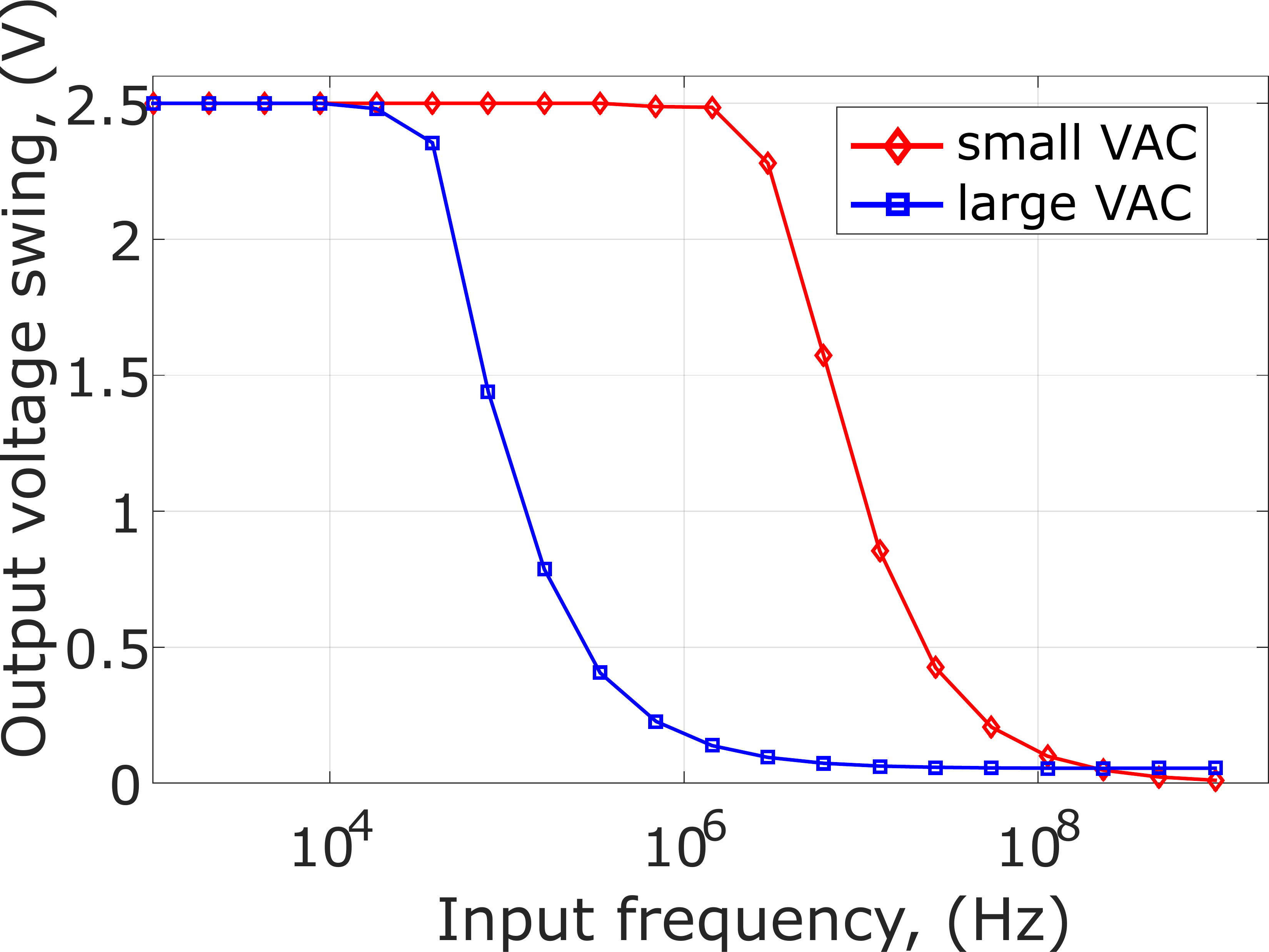}
        \caption{Output voltage swing vs frequency.}
        \label{fig:v_sw_vs_freq}
    \end{minipage} 
\end{figure}

Further simulation experiments are carried out to investigate the VAC's resilience to static frequency variations. Two sizes of the $3 \times 3$ VAC are investigated: the small - with the output capacitor $C_{out}=10pF$ and the output resistors of each cell $R_{out}=100K\Omega$; and the large - with $C_{out}=100pF$ and $R_{out}=1M\Omega$. The duty cycle of all the inputs is $50\%$, and all the weights equal to $7$ (all the cells are enabled). Figure~\ref{fig:v_vs_freq} shows that both VACs produce the output $1.25V$, that equals to $V_{dd}/2$. The average output voltage remains the same on the simulated range of frequencies: from $1kHz$ to $1GHz$.

On the other hand, the value of $C_{out}$ does affect other aspects of perceptron performance. $C_{out}$ contributes to the RC time constant of the VAC circuit, providing a low-pass filter effect on the voltage $DC_{sum}$. As a result, a larger $C_{out}$ is less suitable than a smaller $C_{out}$ for fast response, but would provide better robustness in the presence of frequency variations. In addition, as the voltage to PWM converter depends on the charge on $C_{out}$ for energy, a smaller $C_{out}$ may encounter difficulties in keeping $DC_{sum}$ constant enough to complete the conversion.  

Figure~\ref{fig:v_sw_vs_freq} shows the $DC_{sum}$ voltage swing in the presence of static frequency variations. As can be seen, with reduced input frequency the voltage swing increases, and at some point the VAC operates as a simple inverter with the output voltage $DC_{sum}$ oscillating between $V_{dd}$ and $GND$. Ideally we would like the voltage swing to be not larger than $0.2V$. It means the the frequency of the input PWM signals should not be lower than $1MHz$ for the large VAC and $100MHz$ for the small VAC. 

\begin{figure}[h]

    \centering
    \includegraphics[width=0.45\textwidth]{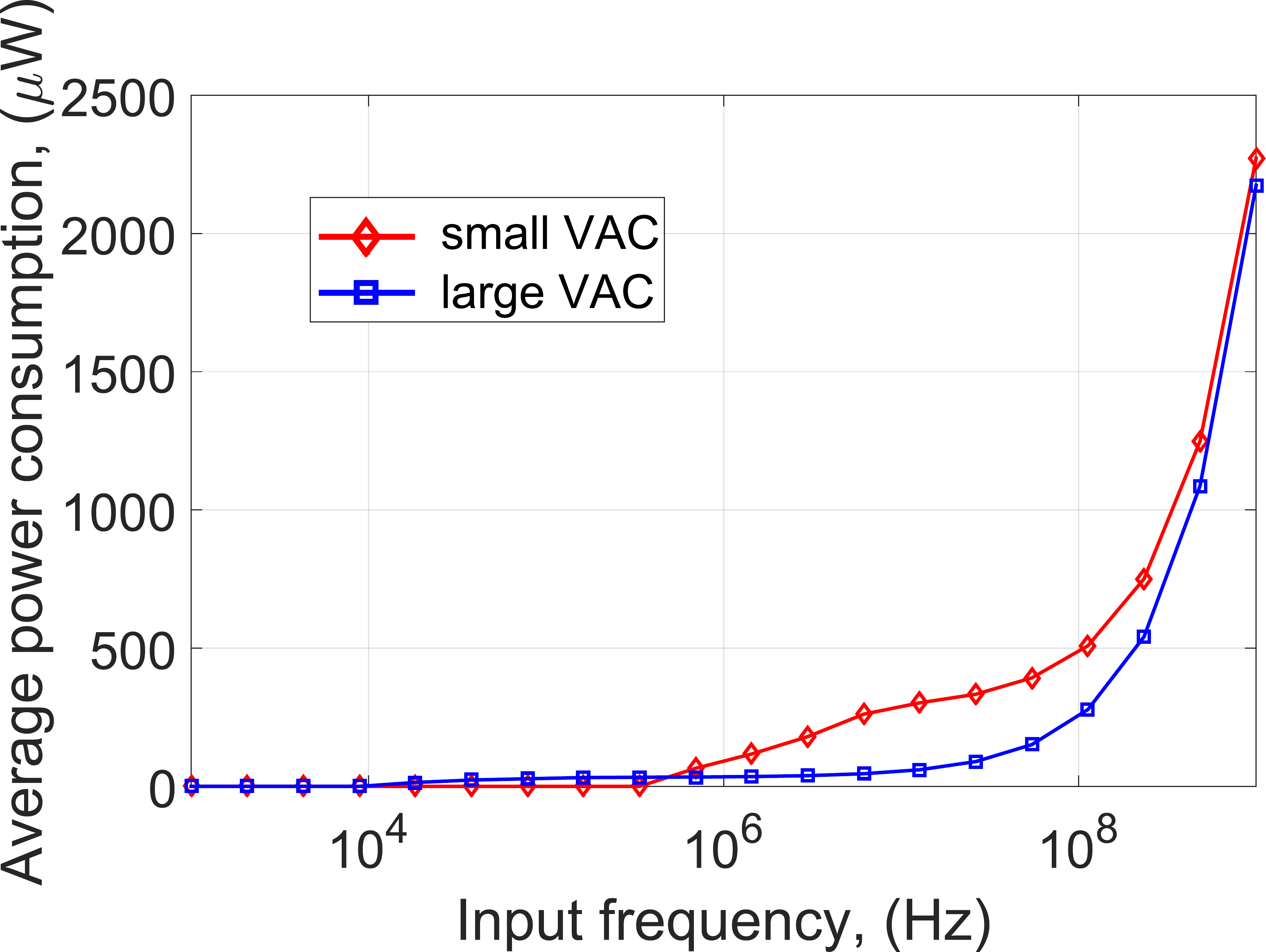}
    \caption{Power vs frequency of the 3x3 VAC.}
    \label{fig:3x3_power_vs_freq}

\end{figure}

In addition, Figure~\ref{fig:3x3_power_vs_freq} shows that VAC size and frequency also affect power consumption. The small VAC has higher power consumption. This is due to the output resistor limiting the charging current. The resistor is 10$\times$ larger in the large VAC, and the current and the power consumption are smaller.

In the large VAC we increase the size of $C_{out}$ and reduce the charging current. This increases the charging time of the capacitor. To investigate this we simulated the time when the voltage on $C_{out}$ reaches the average output value (which is $V_{dd}/2=1.25V$ for the $50\%$ input duty cycle). The capacitor is initially charged to $V_{dd}=2.5V$. The charging time of the capacitor is around $0.14\mu s$ for the small VAC and $14.5\mu s$ for the large VAC, which is true for the entire range of frequencies. This  $\sim 100\times$ ratio is because the $RC$ product is 100$\times$ as large for the large VAC as for the small VAC.

\begin{figure}[h]

    \centering
    \includegraphics[width=0.9\textwidth]{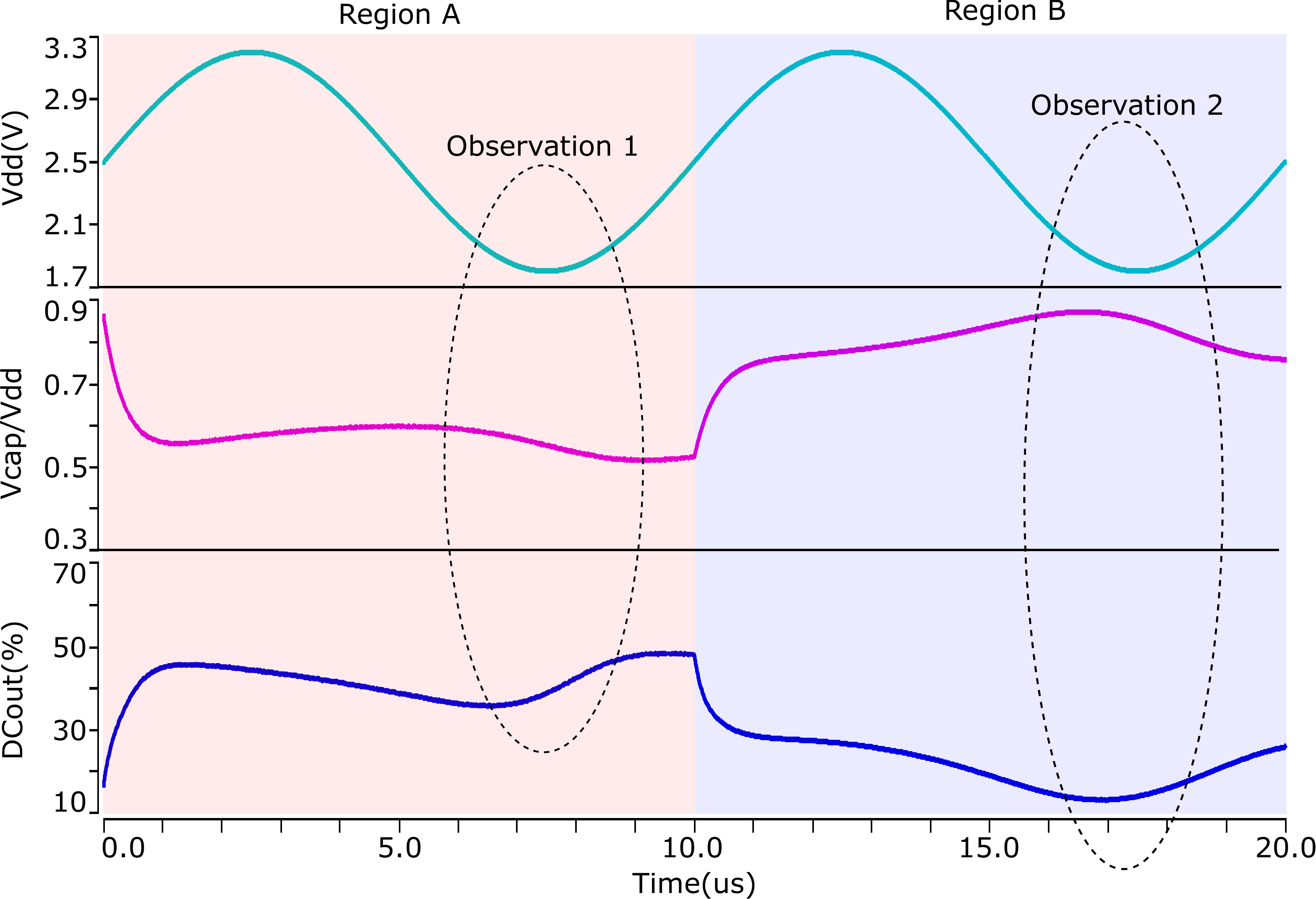}
    \caption{The operation of the perceptron with dynamic power supply voltage variations using an AC supply with 100kHz frequency (for illustration purposes only).}
    \label{fig:voltage_dynamics}

\end{figure}

Figure~\ref{fig:voltage_dynamics} shows the operation of the perceptron with dynamic supply voltage variations to investigate the dynamic power elasticity. The simulations have the following parameters: the cell output resistance $R_{out}=100K\Omega$; the VAC output capacitor $C_{out}=100pF$; the supply voltage varies from $1.8V$ to $3.2V$ with a period of $10\mu s$; input duty cycle $DC_1=DC_2=DC_3=50\%$; the weights are $W_1=W_2=W_3=7$ in region A, and then change to $W_1=W_2=W_3=2$ in region B.

This simulation illustrates the behaviour of the perceptron under very rapid voltage variations at 100kHz frequency (for illustration purposes only). The time period of this change is the same value as the system time constant $RC=10\mu s$. The voltage value swing is also very large - a variation amplitude of 1.4V for a voltage whose nominal value is 2.5V (also used for demonstrating extreme variations). Even under these extreme conditions $V_{cap}/V_{dd}$ still maintains a high degree of resilience. After putting the VAC together with the voltage to PWM converter, however, the combination fares less well, with the output duty cycle changing up to 47\% in Region B (Observation 2). This may be mitigated by improving the compensation mechanism in the voltage to PWM conversion circuit (see Section 2) or by in-situ voltage regulation (not discussed in this paper for brevity).

\subsection {Validation and Analysis of PWM-coded Neural Network}
\label{subsec:pwm_nn_sim_result}
This section explores an NN system built using the proposed PWM-based perceptron. This NN is designed for solving the MNIST problem and has the structure shown in Figure~\ref{fig:nn}. Firstly a high-level model of the perceptron is constructed so that analysis can be carried out in MATLAB, at a higher level than analogue VLSI simulations, which is impractical for systems of this size. Then this model is used in MATLAB investigations on system properties to validate our NN-design approach. 

\paragraph{1. PWM Model and Voltage to PWM Converter Serving as ReLU AF}
\label{subsubsec:pwm_model}
In this section, the duty cycle output of the perceptron is modelled in the form of a mathematical equation with parameters. Then, the model and the voltage to PWM converter itself are studied to verify that the device approximately incorporates the capped ReLU AF.

The equations in Section~\ref{sec:circuit_design}\ref{subsec:pwm_based_nn} pertain to ideal cases. These can be used for comparing with how the implemented perceptron actually delivers. In order to make this comparison at the whole system level, we need to generate a high-level mathematical model based on observations made whilst experimenting with the perceptron circuit at low level.

We experimented in the Cadence Analog Design Environment with a single perceptron, two perceptrons connected in series, mimicking the simplest two-layer NN, and three perceptrons connected in series, emulating the simplest NN with a depth of 3. This is as far as analogue VLSI simulations could practically go, as the three-layer study took many hours on a competitively specified server machine. 

\begin{figure}[h]
    \centering
    \begin{minipage}[b]{0.45\linewidth}
        \centering
        \includegraphics[width=\textwidth]{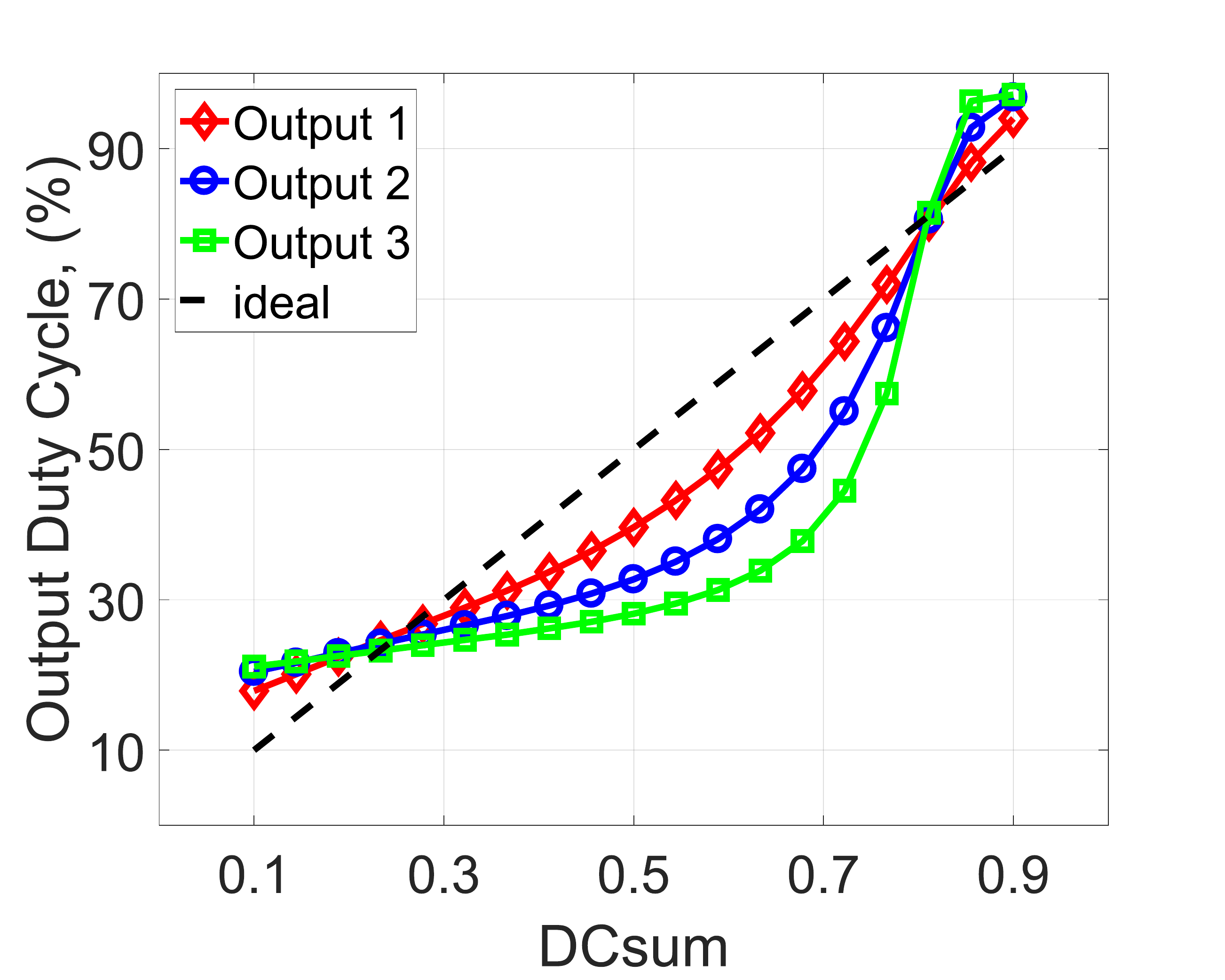}
        \caption{PWM output.}
        \label{fig:pwm_output}
    \end{minipage}
    \hspace{0.5cm}
    \begin{minipage}[b]{0.45\linewidth}
        \centering
        \includegraphics[width=\textwidth]{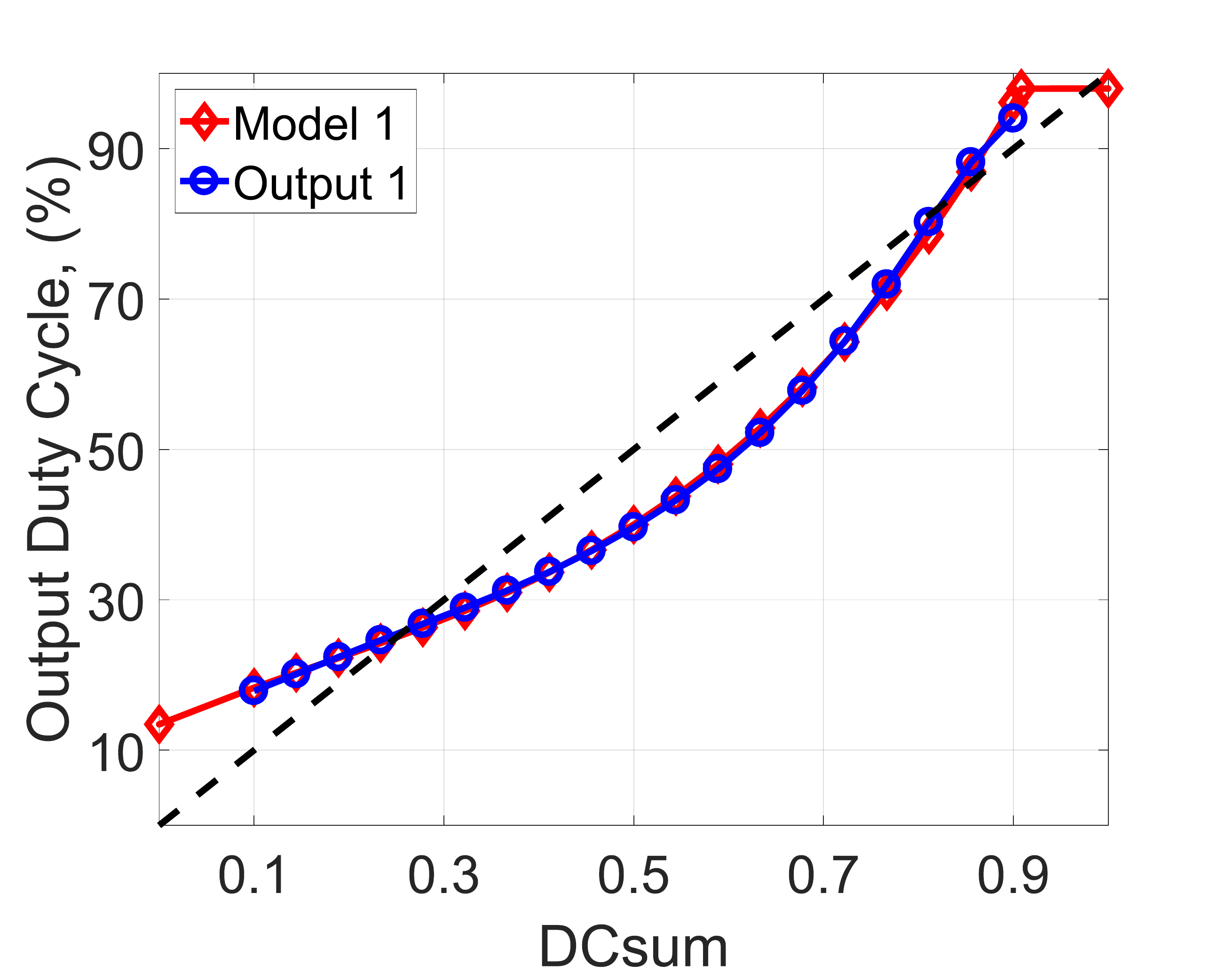}
        \caption{Output vs model stage 1.}
        \label{fig:output_vs_model_1}
    \end{minipage}
\end{figure}

The outputs of these perceptron connection topologies are shown in Figure~\ref{fig:pwm_output}.
In the ideal case, the input and output duty cycles should be equal when every weight is at the maximum value (dashed line).
However, there is a non-linear relationship between the input and output of the single perceptron (red line) and the degree of non-linearity increases when the depth of the NN is increased (blue and green lines).
In addition, the output begins to saturate in the last (third) stage when the input ($DC_{sum}$) reaches 0.85. 

\begin{figure}[h]
    \centering
    \begin{minipage}[b]{0.45\linewidth}
        \centering
        \includegraphics[width=\textwidth]{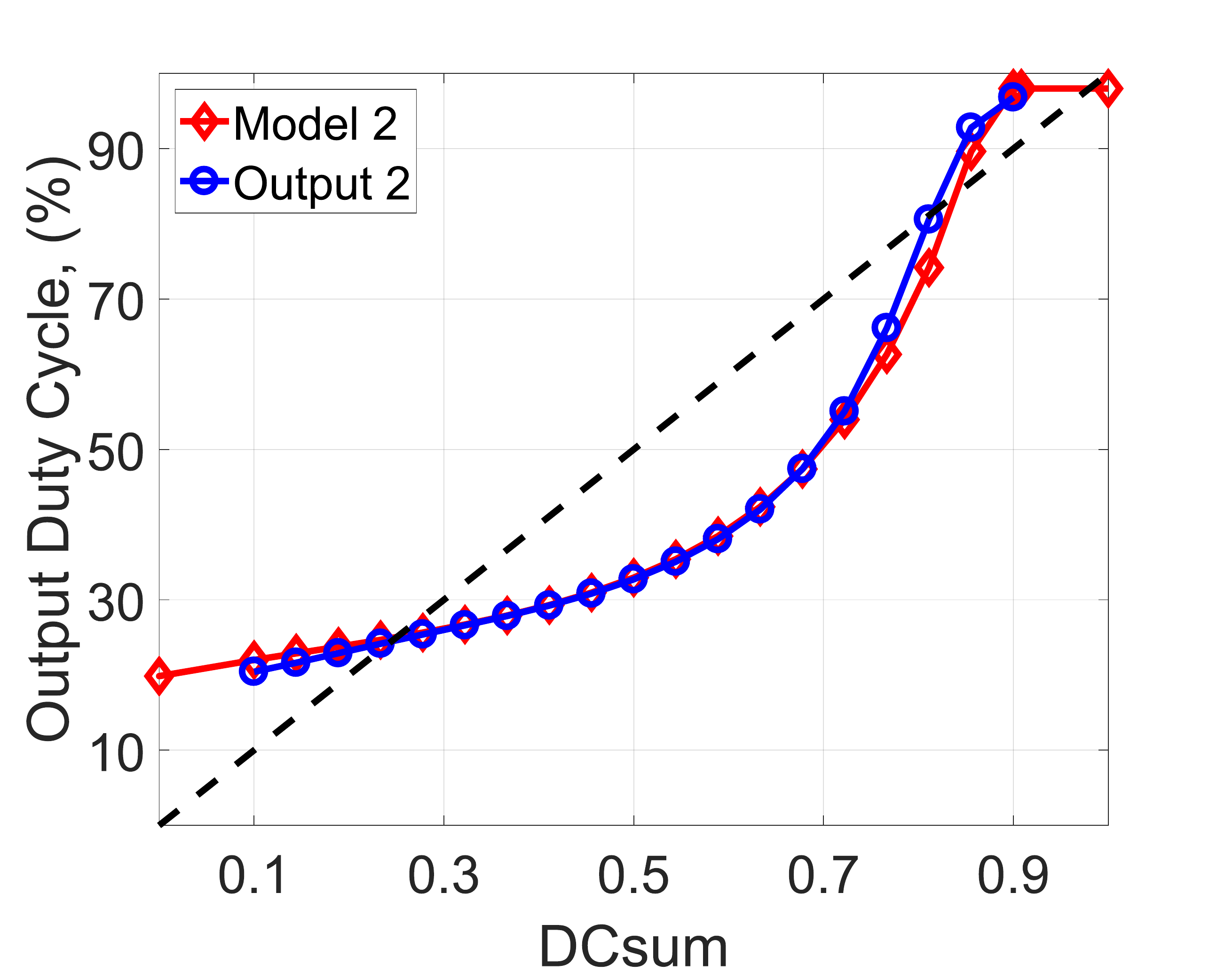}
        \caption{Output vs model stage 2.}
        \label{fig:output_vs_model_2}
    \end{minipage}
    \hspace{0.5cm}
    \begin{minipage}[b]{0.45\linewidth}
        \centering
        \includegraphics[width=\textwidth]{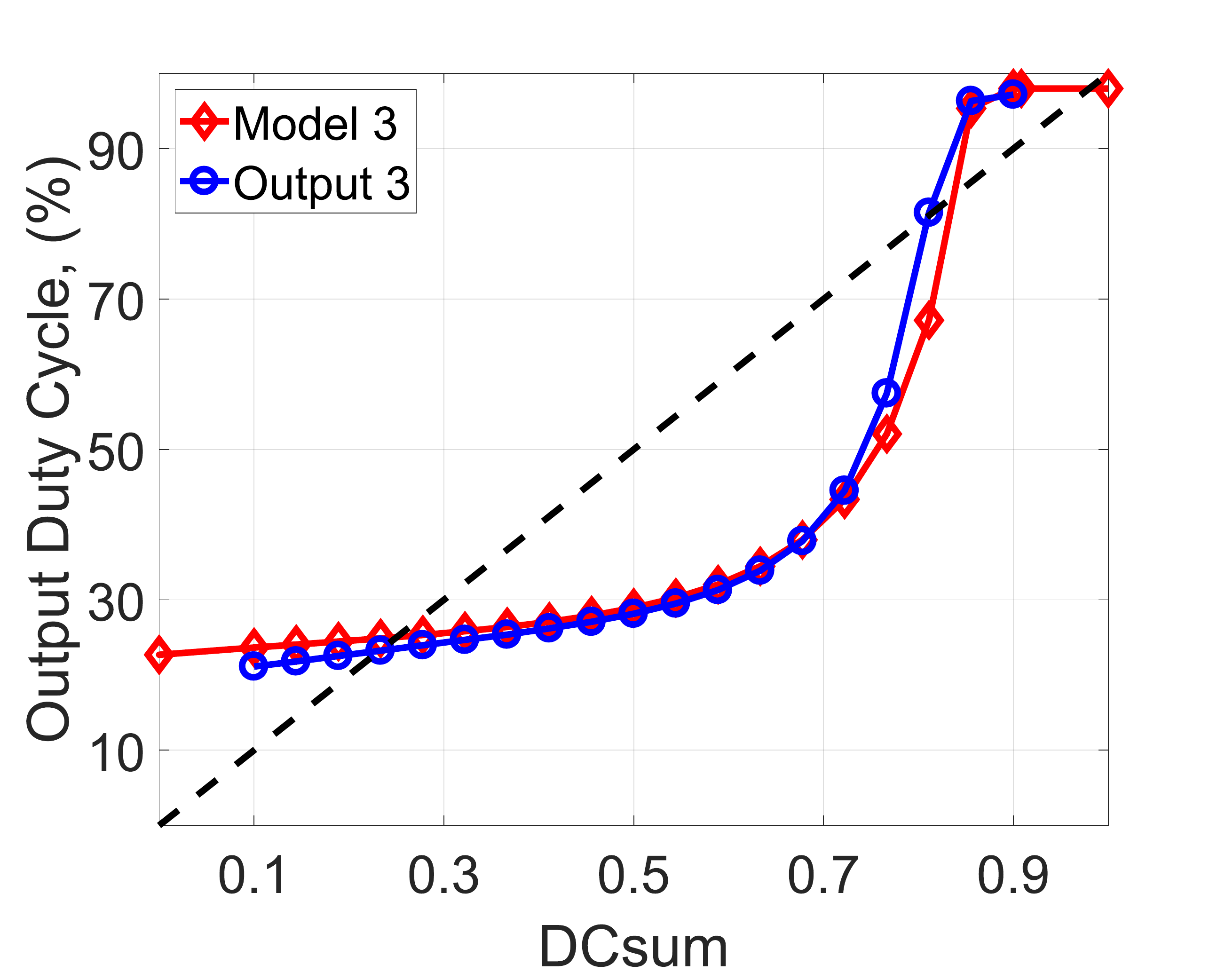}
        \caption{Output vs model stage 3.}
        \label{fig:output_vs_model_3}
    \end{minipage}
\end{figure}

To model this relationship, a third-order polynomial equation, which is easy to differentiate, is curve-fitted to the response of the single perceptron using basic regression in MATLAB. The result is shown in equation~\ref{eq:model}.
Note that the saturation point of the model is set at the maximum output duty cycle, which is $98\%$.
Then, the model is connected in the same two- and three-stage series topologies as in the Cadence experiments and their outputs are plotted in Figure~\ref{fig:output_vs_model_1}~-~\ref{fig:output_vs_model_3}, together with the relevant Cadence results for comparison.
All figures show that this model accurately estimates the input-output relationship of the perceptron. The accuracy can be obtained as the R-squared values of stages one, two and three, which are $99.88\%$, $99.33\%$ and $97.66\%$ respectively.

\begin{equation}
\label{eq:model}
	DC_{out} = 107.27V_{Cout}^3 -53.25V_{Cout}^2 + 52.92V_{Cout} + 13.44
\end{equation}
\begin{equation}
\label{eq:pwm_relu}
f(x)=
\begin{cases}
0 &, x<0
\\
DC_{out} &, 0<x<1
\\
0.98 &, x>0.98
\end{cases}
\end{equation}

\begin{figure}[h]
    \centering
    \begin{minipage}[t]{0.45\linewidth}
        \centering
        \includegraphics[width=\textwidth]{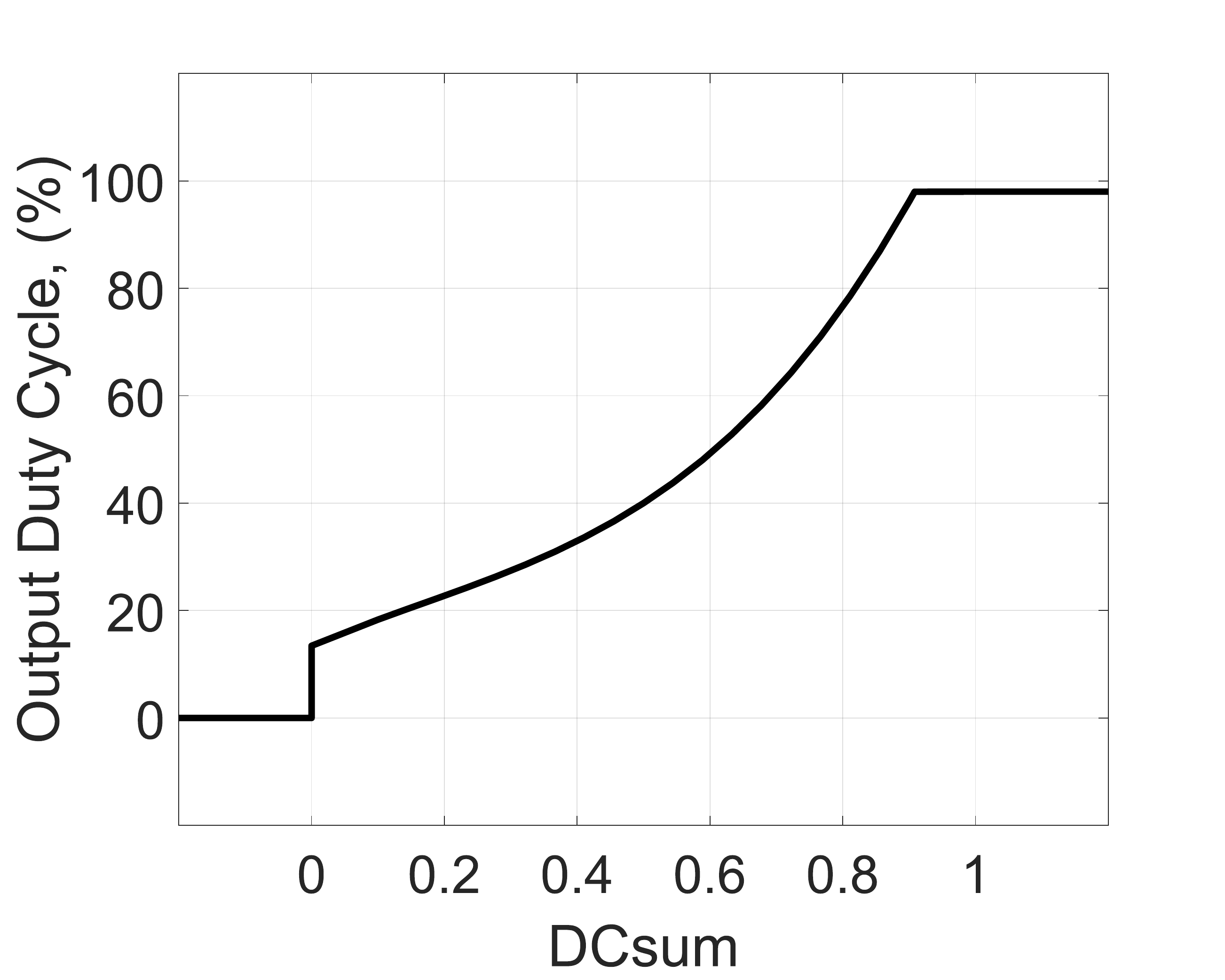}
        \caption{PWM perceptron output model function.}
        \label{fig:relu_pwm}
    \end{minipage}
    \hspace{0.5cm}
    \begin{minipage}[t]{0.45\linewidth}
        \centering
        \includegraphics[width=\textwidth]{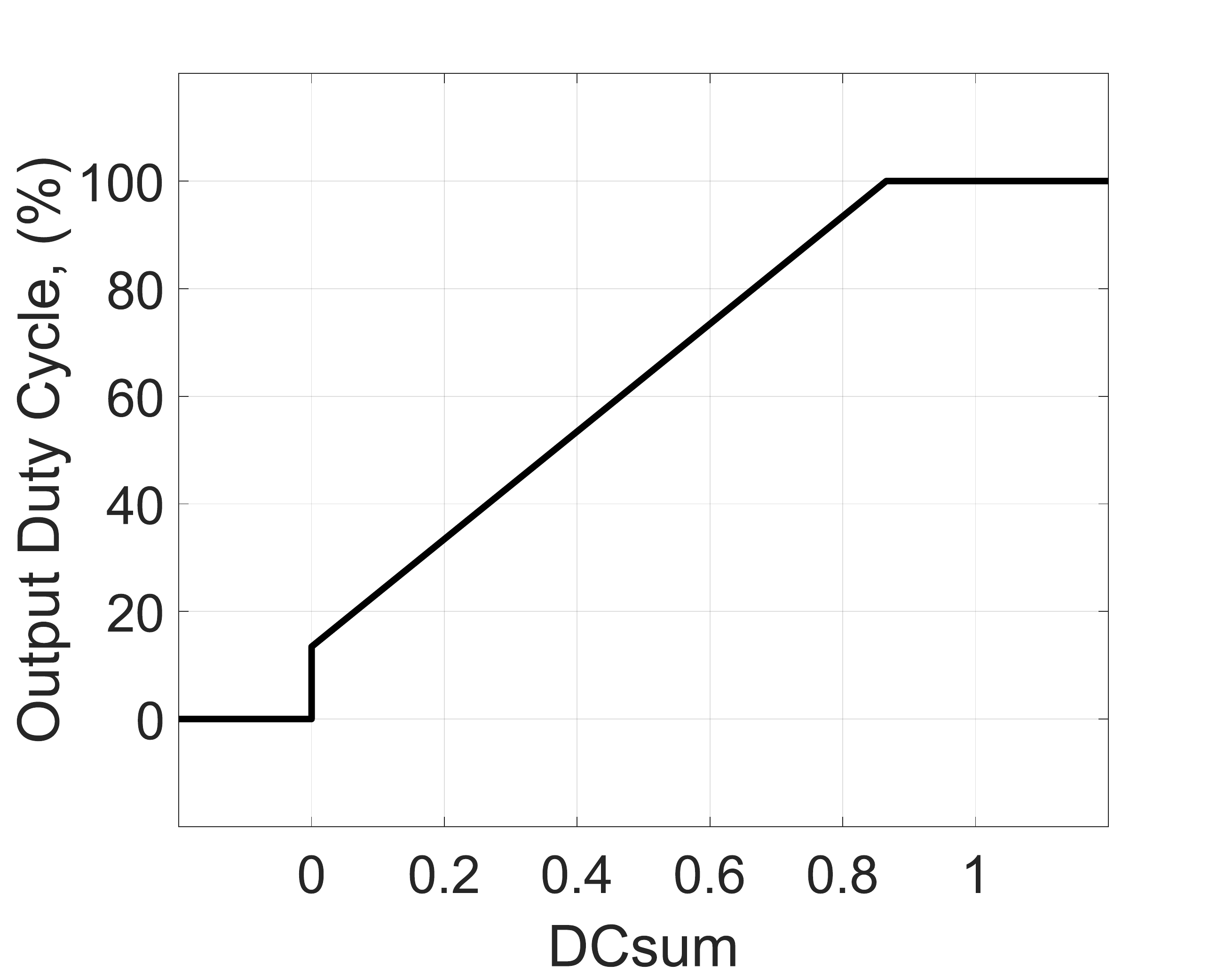}
        \caption{Capped ReLU function with PWM-like offset.}
        \label{fig:relu_offset}
    \end{minipage}
\end{figure}

\paragraph{2. PWM-based NN Simulations}
\label{subsubsec:nn_simumations}

We use the perceptron model in equation~\ref{eq:model} to assemble models of large-size MNIST NNs then simulate these systems in full in MATLAB. The model plot is shown in Fig.~\ref{fig:relu_pwm}.
We also create a capped ReLU function with offset (Oft.ReLU), expressed in equation~\ref{eq:model_relu_offset}. 
As can be seen in the equation, this offset ReLU function takes the constant 13.44 from the perceptron model in equation~\ref{eq:model} to have the same offset nonlinearity as the perceptron model in Fig.~\ref{fig:relu_pwm}, but otherwise has a similar straight line behaviour to the non-offset capped ReLU in Fig.~\ref{fig:relu_capped}. The plot of this offset ReLU can be found in Fig.~\ref{fig:relu_offset}. 

This function is used to investigate whether the step nonlinearity or the curvature nonlinearity higher in the curve of equation~\ref{eq:model} is more important when it comes to NN performance, through comparisons with both the perceptron model in equation~\ref{eq:model} and the capped ReLU function without offset in Fig.~\ref{fig:relu_offset}. 

\begin{equation}
\label{eq:model_relu_offset}
f(x)=
\begin{cases}
0 &, x<0
\\
DC_{out}+13.44 &, 0<x<1
\\
1 &, x>0.86.56
\end{cases}
\end{equation}

There are two groups of simulations using MATLAB: without/with limiting the maximum weight.
All implement the training procedure described in Figure~\ref{fig:nn_training} for the MNIST problem, which is selected as our benchmark.
Without defining the maximum weight, the weight is adjusted freely like the basic FP training while the proposed NN is demonstrated by the limited weight simulation.
Four AFs: ReLU, capped ReLU (Cap.ReLU), capped ReLU with offset (Oft.ReLU) and PWM perceptron (PWM percept.) are applied in three network configurations: two (784/10), three (784/300/10) and four (784/300/100/10) layers. 
The PWM perceptron AF is implemented by the PWM perceptron on its own unmodified - the justification being that it may be considered as an approximation of the capped ReLU (cf. Figure~\ref{fig:relu_pwm} and Figure~\ref{fig:relu_capped}).

As can be seen from this data, these systems being simulated include hundreds of perceptrons and are well beyond analysing in the VLSI design domain. 

For the unlimited weight simulation, the learning rate is swept from 0.001 to 0.1 for every AF. 
The configurations with the smallest error are listed in in Table~\ref{tab:result_float_weight}.
The limited weight simulation is carried out  in the same way except that the initial weight is swept from $\pm 1$ to $\pm 255$.
This is because that a small weight causes a small update which can keep the final weight within the specified range.
Then, the configurations with higher than 90\% accuracy are selected to sweep their maximum weights down until the accuracy is nearly equal to 90\%.
This is to save the circuit area by using the smallest bit-width.
The simulation results are listed in Table~\ref{tab:result_int_weight}.

\begin{table*}[t]
	\centering
	\caption{Simulation result of the floating-point weight neural network.}
	\label{tab:result_float_weight}
	\begin{tabular}{|r|c|c|c|r|c|r|}
		\hline
		\multirow{2}{*}{\textbf{No.}} & 
		\textbf{No.} & 
		\textbf{Activation} & 
		\textbf{Learning}  & 
		\multirow{2}{*}{\textbf{Error}}
		\\ 
		&
		\textbf{Perceptron} &
		\textbf{Function} &
		\textbf{Rate} &
		\\
		\hline \hline
		1 & 784/10 &            ReLU            & 0.010 & 1.40   \\ \hline
		2 & 784/10 &            Cap.ReLU        & 0.008 & 1.75   \\ \hline
		3 & 784/10 &            Oft.ReLU        & 0.009 & 5.09   \\ \hline
		4 & 784/10 &            PWM percept.    & 0.004 & 8.54   \\ \hline \hline
		5 & 784/300/10  &       ReLU            & 0.040 & 1.63   \\ \hline
		6 & 784/300/10  &       Cap.ReLU        & 0.009 & 1.91   \\ \hline
		7 & 784/300/10 &            Oft.ReLU        & 0.002 & 79.54   \\ \hline
		8& 784/300/10  &       PWM percept.    & 0.004 & 27.01   \\ \hline \hline
		9& 784/300/100/10 &    ReLU            & 0.040 & 2.07  \\ \hline
		10& 784/300/100/10 &    Cap.ReLU        & 0.010 & 3.60  \\ \hline
		11 & 784/300/100/10 &            Oft.ReLU        & 0.010 & 90.20   \\ \hline
		12& 784/300/100/10 &    PWM percept.    & 0.090 & 79.07  \\ \hline
	\end{tabular}	
\end{table*}
\begin{table*}[t]
	\centering
	\caption{Simulation result of the integer weight neural network.}
	\label{tab:result_int_weight}
	\begin{tabular}{|r|c|c|c|r|r|r|}
		\hline
		\multirow{2}{*}{\textbf{No.}} & 
		\textbf{No.} & 
		\textbf{Activation} & 
		\textbf{Learning}  & 
		\textbf{Initial} &
		\textbf{Max} &
		\multirow{2}{*}{\textbf{Error}}
		\\ 
		&
		\textbf{Perceptron} &
		\textbf{Function} &
		\textbf{Rate} &
		\textbf{Weight}&
		\textbf{Weight}&
		\\
		\hline \hline
		1 & 784/10          & ReLU          & 0.030 & $\pm$3 & $\pm$31  & 9.28   \\ \hline 
		2 & 784/10          & Cap.ReLU      & 0.040 & $\pm$3 & $\pm$63   & 6.12   \\ \hline 
		3 & 784/10          & Oft.ReLU      & 0.004 & $\pm$7 & $\pm$255  & 7.10 \\ 
		\hline  
		4 & 784/10          & PWM percept.  & 0.030 & $\pm$1 & $\pm$255  & 9.98   \\ \hline 
		\hline
		5 & 784/300/10      & ReLU          & 0.020 & $\pm$255 & $\pm$255  & 79.49  \\ \hline
		6 & 784/300/10      & Cap.ReLU      & 0.020 & $\pm$255& $\pm$255  & 79.49  \\ \hline
		7 & 784/300/10      & Oft.ReLU      & 0.020 & $\pm$3 & $\pm$255  &  18.35 \\ \hline
		8 & 784/300/10     & PWM percept.  & 0.010 & $\pm$15  & $\pm$255  & 25.17  \\ \hline 
 		\hline 
		9 & 784/300/100/10 & ReLU          & 0.010 & $\pm$31 & $\pm$255  & 88.50  \\ \hline
		10 & 784/300/100/10 & Cap.ReLU      & 0.010 & $\pm$31 & $\pm$255  & 88.50  \\ \hline
		11 & 784/300/100/10 & Oft.ReLU      & 0.020 & $\pm$127 & $\pm$255  & 64.09  \\ \hline
		12 & 784/300/100/10 & PWM percept.  & 0.010 & $\pm$63& $\pm$255  & 53.25  \\ \hline
	\end{tabular}	
\end{table*}

\section{Discussions}

This section discusses the results, and highlights the challenges and opportunities in the proposed approach. The section is organised hierarchically from perceptron circuit design to NN experiments validating through the MNIST benchmark application. Towards the end of the section, we relate to our original hypothesis and summarise the key features of the proposed design approach.

\label{sec:discussions}

\subsection{PWM-coded Perceptron}
\label{sec:perc_discussion}

The design of the PWM-based VAC is shown in Figure~\ref{fig:out_vs_vdd_abs} to have satisfied its main design aim, which is to provide resilience in the presence of power supply uncertainty. Adopting a PWM-based approach in order to transfer computation from the digital domain to the relative temporal domain resulted in a device which is essentially independent of the value of the supply $V_{dd}$.

The use of an analogue voltage across a capacitor to represent the result of perceptron arithmetic computations allowed the use of the simplest digital gate, the inverter, to be the fundamental building block for both parts of the perceptron. Programmability, required for the FP functionality of NNs, for instance, can then be realised by using the next simplest digital gate, a two-input NAND gate. This results in an approach which implements digital computations using the smallest digital components working in the relative temporal domain on analogue values. Trading potential loss of precision for power resilience in this way is acceptable for NN applications, as they tend to be accuracy resilient at the point of any particular perceptron. The reduction of circuit complexity and avoidance of conventional multipliers and adders should also contribute to savings in both circuit size and energy consumption. 

Both parts of the perceptron have been shown to have acceptable quality, including being reasonably linear within their operating ranges. However, after putting both together, the perceptron as a whole has certain range problems because the VAC output voltage may be outside the linear region of transistors in the voltage to PWM converter, which is a design assumption for that part. A single-transistor glue logic is then shown to help mitigate this problem. 

The entire perceptron, however, shows weaker power supply variation resilience than the VAC on its own, under extreme dynamic $V_{dd}$ variation conditions. This is primarily because of the following two factors:

\begin{itemize}
    \item as can be seen from Figure~\ref{fig:v_to_pwm}, the oscillator used to convert the voltage on the capacitor between the two parts of the perceptron ($DC_{sum}$) to PWM ($DC_{out}$) draws different amounts of power from the capacitor under different values of $V_{dd}$, and
    \item the simple glue logic in the form of a PMOS transistor does not fully compensate for this because its main function is to keep the current going once the voltage across the capacitor $DC_{sum}$ gets down to threshold.
\end{itemize}

 Computationally, if the correct value of $DC_{sum}$ should be below threshold, which is entirely possible coming from the VAC, this glue logic will cause an error in that value by keeping $DC_{sum}$ at threshold, leading to $DC_{out}$ becoming inaccurate. This is preferable to having the voltage to PWM part dying but computation correctness is still lost.

The above discussion is supported by observing the behaviour shown in Figure~\ref{fig:voltage_dynamics}. The most likely places for errors to appear are when $V_{dd}$ becomes low. In only some of these cases (Observation 2 but not Observation 1) the computation in the VAC would lead to a low $DC_{sum}$ value which might dip below threshold. This will cause the output duty cycle $DC_{out}$ to become incorrect. During Observation 1, although $V_{dd}$ dips low, the computation because of the weights etc. produces a high relative value result that manages to keep $DC_{sum}$ above threshold. This is not the case during Observation 2. On the other hand, if $V_{dd}$ itself reduces below threshold, no matter what value the VAC produces the voltage to PWM part will produce relatively large errors as the perceptron's glue logic would kick in anyway. 

Also of concern is the accumulation of non-linearity after both parts of the perceptron have been put together, and this non-linearity continues to increase once the perceptron is connected in series across multiple stages of an NN of non-trivial depth. 

Another issue is the fact that there may be difficulties for the perceptron to implement AFs other than flavours of ReLU. Of particular concern is that the perceptron, because of the use of the voltage across a grounded capacitor to represent a crucial value, only works in the positive value domain.

Future research topics include the better matching between the constituent parts of the perceptron to overcome the threshold and non-linearity problems and the extension of the perceptron to cover a larger set of AFs.

\subsection {PWM-coded Neural Network}
\label{subsec:pwm_nn_sim_discuss}
In the single perceptron, two-perceptron and three-perceptron experiments, both the resultant model and the Cadence simulations indicate that the voltage to PWM converter, without modifications, may serve as an approximate capped ReLU AF, qualitatively. In addition, the three-perceptron, three-stage full analogue simulation analysis shows that the single-perceptron MATLAB model can be used in multi-stage system analysis without worrying about the fidelity of high-level MATLAB models when multiple layers of perceptrons exist in a system. 

Quantitatively, however, the use of a nonlinear perceptron to approximate linear behaviour becomes increasingly problematic when the depth of the network increases, as the non-linearity accumulates. This is shown to be true by the subsequent whole-NN experiments.

The unlimited weight simulation result in Table~\ref{tab:result_float_weight} gives us traditional NN examples which contain FP weights.
It shows that both ReLU and capped ReLU functions give less than 4\% errors at every depth. 
The results for the PWM perceptron AF are similar to those from the capped ReLU with offset.
They obtain small error rates at the shallowest NN depth, while the capped ReLU without offset outperforms all others at every depth.
This confirms that it is mainly the step transition at $DC_{sum}=0$ that causes the convergence problem in our NN, more than the curvature nonlinearity higher in the curve of Fig.~\ref{fig:relu_pwm}.
Therefore, compensating the circuit to shift the output duty cycle back to 0\% appears to be a promising route of investigation. This will be a subject in our future work. 

Table~\ref{tab:result_int_weight} shows the results with weight limitations.
All results at two-layer NN are worse than the ones in Table~\ref{tab:result_float_weight} due to the weight capping and rounding, except for the PWM perceptron, which does better. The PWM perceptron continues to perform better at higher layer depths than in the unlimited weight case, confirming an advantage for  it when weight is limited and represented by an integer. However, it again fails to approximate the ReLU function quantitatively at higher layer depths, by returning obviously better performances than the latter.

These results confirm the discussions in the previous section. Our proposed PWM-based perceptron's non-linearity as well as not being able to properly extend to the low-percentage range of the PWM duty cycle (it starts from $\sim$ 13\% rather than 0\% as shown in Figs.~\ref{fig:output_vs_model_1} and~\ref{fig:relu_pwm}) makes it less suitable for deeper NNs. Its lack of support for negative values also limits its wider usability as the fundamental element of NNs, without further modification to better incorporate established AFs. The approximation of ReLU, although qualitatively promising, proves to be quantitatively unsatisfactory at higher NN depths, although in some cases this results in the perceptron's AF being better than the ReLU AF.
The high error rate also comes from the resolution loss in basic weight rounding which may be solved by implementing a rounding technique and PF inference quantization presented in~\cite{Lorenz-2015-arXiv} and~\cite{Jacob-2017-arXiv} respectively. 

In other words, even if the negative value representation problem is solved, computing AFs in the analogue and relative temporal domains remains a challenge that must be solved. As a result, a  future work direction is the incorporation of more general arithmetic operations in these domains, which is needed to improve the accuracy of AF implementations. 

A related and interesting unsolved problem is the 'comparing with target vector' function in Figure~\ref{fig:nn}, which is currently relegated to external controllers. It is a duty cycle in and digital out block and can potentially be designed by extending the methods in this work. 

Table~\ref{tab:performance} summarises our design compared to related work.
The work in~\cite{Shuang-2018-arXiv} quantizes the entire NN and yields the lowest error.
However, it is designed for a digital-based processing unit which contains the CPU-memory bottleneck issue implying extra power budget and latency.
Furthermore, real power measurement is missing as it estimates the power consumption from the literature.
Weight rounding methods are proposed in~\cite{Lorenz-2015-arXiv}. 
Although they achieve the second lowest error, they require binarized input data which is a challenge for analogue applications.
Moreover, it does not include an investigation of hardware implementation.
The memristor crossbar NN in~\cite{Jiang-2018-ISCAS} acquires the lowest accuracy with the highest power consumption even its input image size is reduced. 
Charge trap transistor-based NN which performs MNIST classification from the original data is presented in~\cite{Du-2018-TCAD}.
It mainly aims to save power and requires a specific CMOS technology to fabricate the special transistors. 

\begin{table*}[t]
	\centering
	\caption{Performance comparison.}
	\label{tab:performance}
	\begin{tabular}{|c|c|c|c|c|c|c|c|}
		\hline
		\multirow{2}{*}{\textbf{Work}} & 
		\textbf{Weight} & 
	    \multirow{2}{*}{\textbf{MNIST}} & 
		\textbf{NN}  & 
		\multirow{2}{*}{\textbf{Error}} & 
		\textbf{Power} &
		\textbf{Power} &
		\multirow{2}{*}{\textbf{Hardware}}
		\\ 
		&
		\textbf{type} &	&
		\textbf{conf.} & &
		\textbf{($\mu W$)} & 
		\textbf{elastic} &
		\\
		\hline 
		\hline
		\cite{Shuang-2018-arXiv} & integer & quantized & WAGE & 0.4 & n/a & N & MCU \\ \hline 
		\cite{Lorenz-2015-arXiv} & integer & binarized & MLP & <3\% & n/a & N & n/a \\ \hline 
		\multirow{2}{*}{\cite{Jiang-2018-ISCAS}} & \multirow{2}{*}{n/a} & reduced \& & \multirow{2}{*}{MLP} & \multirow{2}{*}{13.5\%} & 53,000 & \multirow{2}{*}{N} & memristor \\ 
		& & quantized & & & (NN) & & crossbar \\ \hline		\multirow{2}{*}{\cite{Du-2018-TCAD}} & \multirow{2}{*}{fixed-point} & \multirow{2}{*}{original} & \multirow{2}{*}{MLP} & \multirow{2}{*}{$\sim$5\%} & 14,800 & \multirow{2}{*}{N} & \multirow{2}{*}{Spec. CMOS} \\ 
		& & & & & (NN) & & \\ \hline
		\multirow{2}{*}{This} & \multirow{2}{*}{integer} & \multirow{2}{*}{original} & \multirow{2}{*}{MLP} & \multirow{2}{*}{<10\%} & 14-1,080 & \multirow{2}{*}{Y} & \multirow{2}{*}{Std. CMOS} \\ 
		& & & & & (VAC) & & \\ \hline
	\end{tabular}	
\end{table*}

Even though the error of our design is higher, it is still within the same order of magnitude, and we have yet to investigate more sophisticated techniques for compensating the voltage to PWM part to improve the duty cycle coverage and eliminate the step nonlinearity at $DC_{sum}=0$, which promises to reduce error. We also do not investigate beyond standard CMOS technology as that is out of scope for this investigation. Our focus is on tolerating unstable and unpredictable power supply voltages, a necessity for energy autonomous AI devices. In this regard, this solution is unique as existing research in the literature invariably requires stable and known voltages and operating frequencies. 
Note that the power figures in Table~\ref{tab:performance} are not directly comparable. 
The figure for~\cite{Lorenz-2015-arXiv} is obtained from measurements carried out on a fabricated chip at the 28nm technology node with special non-CMOS transistor techniques aiming to showcase the advantage of that technology in low-power operations, whilst that for this work is obtained from simulating one VAC at the 65nm technology node with a deliberately high (2.5V) nominal supply voltage to facilitate studying voltage instability scenarios. 
This is similar to~\cite{Jiang-2018-ISCAS} where the circuit is implemented with non-CMOS technology. 
A fair power consumption comparison with~\cite{Lorenz-2015-arXiv} and~\cite{Jiang-2018-ISCAS} is not yet possible without fabricating and testing real chips, preferably at the same VLSI technology node, as whole-NN simulations at the VLSI level is not practical. In addition, the power figures for~\cite{Lorenz-2015-arXiv} and~\cite{Jiang-2018-ISCAS} are themselves not comparable with each other as~\cite{Jiang-2018-ISCAS} covers the entire system including peripherals and~\cite{Lorenz-2015-arXiv} covers the NN engine only.

\subsection {Overall Summary}
\label{subsec:summary}
Our design methods, supported by extensive analysis and validations, have proven the original hypothesis, and demonstrated the following features:

\begin{itemize}
    \item \textit{power elasticity} and \textit{resilience} across a dynamic range of $V_{dd}$ (statistically varying by 5$\times$) and $f$ (statistically varying by up to 6 orders of magnitude). Such elasticity is achieved for the VAC without requiring any voltage regulator circuit and clock pairing between $V_{dd}$ and $f$. Dynamic power supply variations also show good resilience. However, further compensation will be required at lower voltages to avoid large errors at the whole-perceptron level;  
    \item minimal use of additional analogue (ideally, only passive components) electronics, coupled with low-complexity PWM-coded arithmetic using primarily digital components, making our approach highly power efficient and suitable for low-cost fabrication;
    
    \item extensive validation and analysis using multi-layer PWM-coded NNs show good scalability of the proposed approach; however, deeper NNs may need circuit-level compensation after each layer or high-precision representation techniques to improve the overall accuracy and efficiency.
\end{itemize}

\section{Conclusions}

\label{sec:concl}

We propose the first mixed-signal (analogue\slash digital\slash relative temporal) perceptron design using the principles of PWM. Central to our design are a number of parallel inverters that suitably transcode the input-weight pairs from the spatial domain to the relative temporal domain. This approach aims to deliver high resilience to amplitude and frequency variations in the supply voltage, exploiting the fact that PWM-based solutions are typically agnostic to such variations. 

Another advantage of the proposed design is its simplicity. Whilst conventional implementations of the perceptron require complex logic to perform multiplication and addition, the proposed approach uses only one gate (either an inverter or a two-input NAND) per bit for every input. Thus, for the $3 \times 3$ weighted addition VAC we used only 54 transistors. This significantly reduces the logic requirement and, therefore, the power consumption of the entire device.

Extensive experimentation on the perceptron and its use in neural networks of relatively significant sizes helps to explore the perceptron design's advantages, usability and limitations. Also through experimental studies, design improvements are found which further strengthen the perceptron's case. These experimental explorations also lead to further insights into the design and provide guidance on potential future work.

The perceptron's arithmetic unit design is shown to fully accomplish its design aim of power and frequency resilience. It is also shown to be working within reasonable boundaries in pragmatic applications, especially in NNs of limited depth which are nevertheless of significant size. Future improvements should be concentrated on improving its linearity, its threshold voltage independence and its representation of negative values to increase its usability of NNs of greater depth and more sophisticated AFs. 

This work points to potentially exciting research to extend the computation capabilities of devices working in the relative temporal and analogue domains. 

Machine learning is finding more applications at the micro-edge, where power variation from energy harvesters is becoming commonplace. We believe the proposed perceptron will find practical implementations in these applications as it is highly robust to these variations.

\vskip6pt

\enlargethispage{20pt}

\ethics{Insert ethics statement here if applicable.}

\dataccess{Insert details of how to access any supporting data here.}

\aucontribute{Serhii Mileiko was responsible for the low-level perceptron and NN circuit design, experiments and anlysis, and also led the writing of the paper. Thanasin Bunnam was responsible for the scale-up NN circuit modeling, analysis and experiments using MNIST benchmark application, and co-led the writing of the paper. Fei Xia  contributed to the writing of the paper, led its editing, and  participated in technical discussions. Rishad Shafik contributed in the scale-up models, tied up with the low-level circuits, co-supervised the circuit- and system-level works of S. Mileiko and T. Bunnam, and contributed to writing\slash editing. Alex Yakovlev contributed by proposing the idea of using CMOS logic for duty-cycle based computing for power elasticity and robustness, supervising the work of S. Mileiko and T. Bunnam, and all technical discussions. Shidhartha Das has contributed in the circuit design aspects, particularly reflecting on how averaging could affect the overall accuracy.}

\competing{The authors declare that they have no competing interests.}

\funding{This work was supported by the EPSRC STRATA project (EP/N023641/1).}

\ack{The authors would like to thank Mr. Jonathan Edwards, CTO of Temporal Computing Ltd., for his useful suggestions in the early stages.}


\bibliographystyle{IEEEtran}
\bibliography{refs}
\end{document}